\documentclass[aps,prd,reprint,twocolumn,superscriptaddress,longbibliography,nofootinbib,floatfix,showpacs]{revtex4-1}
\usepackage{epsfig}
\usepackage{amsmath,amssymb,amsfonts}
\usepackage{hyperref}
\usepackage{mathrsfs}
\usepackage{bbm}
\usepackage{slashed}
\usepackage{graphicx}
\usepackage{verbatim}

\usepackage{bm}

\usepackage{bbm}

\usepackage{rotating}

\usepackage[usenames]{color}

\definecolor{darkgreen}{rgb}{0.2,0.6,0}

\newcommand{\be}{\begin{equation}}
\newcommand{\ee}{\end{equation}}
\newcommand{\bw}{\begin{widetext}}
\newcommand{\ew}{\end{widetext}}
\newcommand{\bi}{\begin{itemize}}
\newcommand{\ei}{\end{itemize}}
\newcommand{\bea}{\begin{eqnarray}}
\newcommand{\eea}{\end{eqnarray}}
\newcommand{\ud}{\mathrm{d}}

\newcommand{\LCm}{{\scriptscriptstyle -}} 
\newcommand{\LCp}{{\scriptscriptstyle +}}
\newcommand{\LCpm}{{\scriptscriptstyle \pm}}
\newcommand{\LCmp}{{\scriptscriptstyle \mp}}

\newcommand{\LCperp}{{\scriptscriptstyle \perp}}

\usepackage[T1]{fontenc} \usepackage[latin1]{inputenc}

\begin{document}
%

\title{Nonlinear trident in the high-energy limit: Nonlocality, Coulomb field and resummations}

\author{Greger Torgrimsson}
\email{g.torgrimsson@hzdr.de}
\affiliation{Helmholtz-Zentrum Dresden-Rossendorf, Bautzner Landstra{\ss}e 400, 01328 Dresden, Germany}
\affiliation{Theoretisch-Physikalisches Institut, Abbe Center of Photonics,
Friedrich-Schiller-Universit\"at Jena, Max-Wien-Platz 1, D-07743 Jena, Germany}
\affiliation{Helmholtz Institute Jena, Fr\"obelstieg 3, D-07743 Jena, Germany}

\begin{abstract}
We study nonlinear trident in laser pulses in the high-energy limit, where the initial electron experiences, in its rest frame, an electromagnetic field strength above Schwinger's critical field.
At lower energies the dominant contribution comes from the ``two-step'' part, but in the high-energy limit the dominant contribution comes instead from the one-step term. We obtain new approximations that explain the relation between the high-energy limit of trident and pair production by a Coulomb field, as well as the role of the Weizs\"acker-Williams approximation and why it does not agree with the high-$\chi$ limit of the locally-constant-field approximation. We also show that the next-to-leading order in the large-$a_0$ expansion is, in the high-energy limit, nonlocal and is numerically very important even for quite large $a_0$. We show that the small-$a_0$ perturbation series has a finite radius of convergence, but using Pad\'e-conformal methods we obtain resummations that go beyond the radius of convergence and have a large numerical overlap with the large-$a_0$ approximation. We use Borel-Pad\'e-conformal methods to resum the small-$\chi$ expansion and obtain a high precision up to very large $\chi$. We also use newer resummation methods based on hypergeometric/Meijer-G and confluent hypergeometric functions.  
\end{abstract}
\maketitle

\section{Introduction}

Quantum electrodynamics in strong laser fields is usually studied by treating the interaction with the quantized photon field in a standard perturbation expansion in $\alpha=e^2/(4\pi)$, but with a Volkov/Furry picture treatment of the strong field. The strength of the field is usually described in terms of the ``classical nonlinearity parameter'' $a_0=E/\omega$\footnote{We use units with electron mass $m_e=1$, and a factor of the charge $e$ has been absorbed into the field strength $eE\to E$.}, where $E$ is the field strength and $\omega$ a typical frequency scale of the in general pulsed plane wave. 
For $a_0>1$ one can in general not treat the field in a perturbation expansion in $a_0$. However, if $a_0\gg1$ one can make an expansion in $1/a_0$~\cite{Dinu:2017uoj,Ilderton:2018nws}, which corresponds to an approximation where the inhomogeneous field is treated as being locally constant. This is a very useful approximation that allows otherwise very complicated processes to be studied. However, $a_0$ is not the only parameter in the system and so how large $a_0$ has to be for this locally-constant-field (LCF) approximation to be valid depends on the momenta of the particles involved~\cite{Dinu:2015aci,DiPiazza:2017raw,Podszus:2018hnz,Ilderton:2019kqp}. 

A plane wave on its own cannot produce any particles, but a single particle entering a plane wave can, and at high intensity such a seed particle can lead to the production of a large number of particles in cascades~\cite{Bell:2008zzb,Elkina:2010up,Nerush:2010fe}. 
So, consider a single particle with momentum $p_\mu$ that enters the strong field.
In a plane-wave field, the integrated/total probability that this particle decays/produces some other particles (e.g. an electron and a positron in trident) only depends on $a_0$ and a second parameter $b_0=kp$, where $k_\mu$ is the null wave vector of the plane wave ($k^2=0$ and $k_0=\omega$). The coefficients in the LCF expansion in $1/a_0\ll1$ only depend on $b_0$ via the ``quantum nonlinearity parameter'' $\chi=a_0b_0$. (Note that $\chi$ is independent of $\omega$, so the LCF expansion can be seen as a derivative expansion.)

The limit where $\chi$ becomes very large, that is $\alpha\chi^{2/3}\gtrsim1$, is different from what one might expect from the high-energy limit of QED without a strong field, and it has been conjectured that the expansion in $\alpha$ might even break down in this regime~\cite{Narozhnyi69,Ritus70,Ritus72,Morozov:1975uah,MorozovNarozhnyi77,Narozhnyi:1979at,Narozhnyi:1980dc,MorozovNarozhnyiRitus}. This would then be a regime where neither the strong field nor the $\alpha$ dependence can be treated with perturbation theory, i.e. QED would be truly strongly coupled. This conjecture is an old result which has attracted a great deal of interest in the last couple of years~\cite{Fedotov:2016afw,Yakimenko:2018kih,Baumann:2018ovl,Blackburn:2018tsn,Mironov:2020gbi}.   

It has recently been shown~\cite{Podszus:2018hnz,Ilderton:2019kqp} that whether the large $\chi$ limit is reached by having $a_0$ or $b_0$ being the largest parameter leads to fundamentally different results. If $a_0$ is the largest parameter then LCF is good, but if $b_0$ is largest then the probabilities of nonlinear Compton scattering and Breit-Wheeler pair production reduce to the leading perturbative results, i.e. they become proportional to $a_0^2$, and then one has ordinary high-energy scalings without the suggestion of an $\alpha$ expansion break down.

In this paper we study and compare these two ways of reaching high $\chi$ for the trident process~\cite{Dinu:2017uoj,King:2018ibi,Mackenroth:2018smh,Acosta:2019bvh,Krajewska15,Hu:2014ooa,King:2013osa,Ilderton:2010wr,Hu:2010ye,Bamber:1999zt,Ritus:1972nf,Baier,Dinu:2019wdw}, $e^\LCm\to e^\LCm+e^\LCm+e^\LCp$. One motivation for this is that for $\alpha\chi^{2/3}\gtrsim1$ one is interested in whether the $\alpha$ expansion breaks down, so it is natural to ask how the results in~\cite{Podszus:2018hnz,Ilderton:2019kqp} generalize to higher-order processes, and the trident process is such an example. The trident probability is of course not the next-to-leading order correction to be added to the nonlinear Compton or Breit-Wheeler probabilities, but it does correspond to the imaginary part of a loop that is part of the $\alpha$ expansion of the $e^\LCm\to e^\LCm$ amplitude (the mass operator). Moreover, for constant fields this loop gives the dominant contribution at $\mathcal{O}(\alpha^2)$ (the loops giving e.g. double nonlinear Compton scattering~\cite{Morozov:1975uah,Lotstedt:2009zz,Loetstedt:2009zz,Seipt:2012tn,Mackenroth:2012rb,King:2014wfa,Dinu:2018efz,Wistisen:2019pwo} is subdominant), see the review~\cite{Fedotov:2016afw} for a collection of the various loops that have so far been calculated.

Another motivation for this study is phenomenological. Part of the trident process was observed at SLAC two decades ago~\cite{Bamber:1999zt}. Since then there have basically not been any new experiments. But there are now definite plans for new trident experiments at e.g. LUXE~\cite{LUXEparameters,Abramowicz:2019gvx} and FACET-II~\cite{MeurenPresentation,Meuren:2020nbw}. In the old SLAC experiment the laser was relatively weak, i.e. $a_0<1$, and, due to the lack of complete theoretical predictions, a Weizs\"acker-Williams (WW) approximation was used to estimate the trident probability. However, the new experiments will have larger $a_0$ and it has been shown~\cite{King:2013osa} that this WW approximation does not give the same result as the LCF approximation. One might still want to use an approximation such as WW, because LCF only works when $a_0$ is sufficiently large, and so it is not clear how good LCF is for $1<a_0<10$, which is a regime that is relevant for upcoming experiments. WW on the other hand is associated with high energy rather than high intensity, so one might expect that WW could be used in regimes where LCF is not good.
In this paper we show that there is indeed a regime where the WW approximation is good. 

We also compare the high-energy limit of trident with pair production by a Coulomb field in a plane wave. To do this we generalize our results in~\cite{Dinu:2017uoj} to a process where the initial electron is replaced by another particle, e.g. a muon, which has the same charge but different mass. 

Studying trident is also relevant as the first step in cascades, i.e. the production of a large number of particles. For trident one can compare approximations with the exact result as a way of determining in what parameter regimes similar approximations can be used for higher-order processes, for which one cannot compare with the exact result. In this context one may ask how the momentum of the initial particle is distributed among the produced particles. In the emission of a single photon by an electron, the probability has in the high-energy limit a peak where the emitted photon takes away almost all of electron's momentum~\cite{Tamburini:2019tzo,Bulanov:2013cga}. In contrast, for trident, where the emitted photon decays into a pair, one finds that the probability is largest when the initial electron keeps almost all of its momentum and only gives a small fraction to the emitted photon and the produced pair. This low-momentum transfer has an important impact on the behavior of the high-energy limit of trident compared to the first-order processes~\cite{Podszus:2018hnz,Ilderton:2019kqp}.

With this study we are also mapping a part of the parameter space not considered in previous literature. At small $a_0$ the probability is perturbative and scales to leading order as $\mathcal{O}(a_0^2)$, which is a regime that has been studied since the 40's~\cite{Borsellino}. For large $a_0$ the leading order scales as $\mathcal{O}(a_0^2)$~\cite{Ritus:1972nf,Baier}, while the full next-to-leading order $\mathcal{O}(a_0)$ was only calculated recently~\cite{Dinu:2017uoj,King:2018ibi}. In~\cite{Dinu:2017uoj} we also considered the low-energy regime and obtained explicit analytical expressions valid for arbitrary $a_0\gtrsim1$.
In this paper we complement these previous studies by providing new analytical results in the high-energy limit, for arbitrary $a_0$.

The rest of this paper is organized as follows. In Sec.~\ref{Definitions} we provide some basic definitions and the generalization of some of our results in~\cite{Dinu:2017uoj} to muon trident. In Sec.~\ref{LCFsection} we consider the large-$\chi$ limit of the LCF approximation, i.e. we first take $a_0$ to be the largest parameter and then we take $\chi$ large. In Sec.~\ref{highEnergySection} we study the limit where $b_0$ is the largest parameter. We show that having $a_0$ largest and then $b_0$ large does not commute with having $b_0$ largest and then $a_0$ large. In Sec~\ref{CoulombSection} we compare our new high-energy approximation with pair production by a Coulomb field. In Sec.~\ref{WWsection} we study the applicability of the WW approximation.
In Sec.~\ref{nextToLead} and~\ref{nextToLeadShort} we study the next-to-leading order corrections in the large-$a_0$ expansion of the high-energy approximation, and show that these corrections are nonlocal and numerically important.
In Sec.~\ref{perturbationSection} we study the perturbation series around $a_0=0$. We show that there is a finite radius of convergence. Using Pad\'e approximants and a conformal map we find that the coefficients in the perturbation series can be used to obtain a good approximation beyond the radius of convergence and even for large $a_0$. 
In Sec.~\ref{saddlePointSection} we compare with the low-energy regime in the case where the initial particle is much heavier than the produced pair.
In Sec.~\ref{resumchi} we resum the divergent small-$\chi$ expansion and obtain a resummation that has a high precision up to very large $\chi$.
We conclude in Sec.~\ref{conclusionsSection}.

\section{Definitions}\label{Definitions}

We use $v_\LCperp=\{v_1,v_2\}$ and $v^\LCpm= 2v_\LCmp=v^0\pm v^3$.
In terms of the vector potential the field is given by $a_\LCperp(\phi)$, $a_0=a_3=0$, where $\phi=kx=\omega x^\LCp$.
In order to understand the high-energy limit of the trident process, it is useful to consider the process where the initial electron is replaced by a particle with a different mass but with the same charge. This could for example be muon trident~\cite{Baier,Ritus:1972nf,Muller:2009ri} $\mu^\LCm\to \mu^\LCm+e^\LCm+e^\LCp$, but we will keep the mass $\mu$ of the initial particle arbitrary because we are also interested in comparing with the infinite mass limit.
When the initial particle is an electron then we have two identical particles in the final state, so there are two terms on the amplitude level $M=M_{12}-M_{21}$, where one $M_{12}$ is obtained from the other $M_{21}$ by swapping these identical particles. We refer to the cross term $2\text{Re}M_{21}^*M_{12}$ between these two terms as the exchange part of the probability $\mathbb{P}=\mathbb{P}_{\rm dir}+\mathbb{P}_{\rm ex}$. The exchange term is the most difficult to calculate. Indeed, it was for a long time omitted in the literature, even for the simplest case of constant fields. We have shown though that e.g. for short pulses and moderately high intensity ($a_0\sim1$) the exchange term is important.  
The complicated exchange term is of course absent if the initial particle is not an electron. 
If the initial particle is an electron then the exchange term becomes small compared to the direct part of the probability at high enough energies.

The generalization (of the direct terms) to an arbitrary mass $\mu$ is obtained in the same way as in~\cite{Dinu:2017uoj}, we only have to replace the electron mass (which is $m_e=1$ in our units) with the muon mass $\mu$ in some places. When comparing with~\cite{Baier,Ritus:1972nf} it is important to note that we still use units with $m_e=1$, so for an incoming muon we have $\mu\approx207$. In~\cite{Dinu:2017uoj} we had two identical particles in the final state and therefore had to divide the probability by a factor of $2$ to avoid double-counting when summing over momenta and spin. Since we do not have identical particles here, we have an overall factor of $2$ compared to~\cite{Dinu:2017uoj}. In the identical-particle case one has two contributions to the probability, $|M(s_1,s_2)|^2+|M(s_2,s_1)|^2$, which give the same contribution to the total/integrated probability. So, for the total probability, in the (mathematical) limit $\mu\to1$ our results here reduce to the direct terms in~\cite{Dinu:2017uoj} with the same factors of $2$ in the prefactor.
As in~\cite{Dinu:2017uoj,Dinu:2013hsd} we integrate over the transverse momenta of all the final state particles. The longitudinal-momentum spectrum $\mathbb{P}(s)$ is defined via
\be
\mathbb{P}=\int_0^1\ud s_1\ud s_2\theta(s_3)\mathbb{P}(s) \;.
\ee 
These longitudinal-momentum variables are the ratios $s_i=kp_i/kp$, $s_3=1-s_1-s_2$, and we use $b_0=kp$ for the initial particle. We use $q_1=1-s_1$ for the longitudinal momentum of the intermediate photon. 
In our approach we find it convenient to separate the total probability into three terms,
\be
\mathbb{P}=\mathbb{P}_{11}+\mathbb{P}_{12}+\mathbb{P}_{22} \;,
\ee   
which have different number of lightfront time integrals. There is nothing fundamental about this particular separation. In fact, for a constant field, or for large $a_0$, $\mathbb{P}_{22}$ gives one term that is quadratic in the volume and another term that is linear in the volume, where the latter should then be combined with $\mathbb{P}_{11}$ and $\mathbb{P}_{12}$, which are also linear in the volume. 

Since the calculation is basically the same as in~\cite{Dinu:2017uoj}, we simply state the final results, which are valid for arbitrary field shape and polarization. The simplest term comes from the absolute square of a ``lightfront-time-instantaneous'' term on the amplitude level
\be\label{P11GenFin}
\mathbb{P}_{11}(s)=\frac{2\alpha^2}{\pi^2}\int\!\ud\phi_{12}\frac{1}{q_1^4}\frac{-s_0s_1s_2s_3}{(\theta_{21}+i\epsilon)^2}e^{\frac{i}{2b_0}\left[r_1\Theta_{21}^{\mu}+r_2\Theta_{21}^{e}\right]} \;,
\ee
where $r_1=(1/s_1)-(1/s_0)$, $r_2=(1/s_2)+(1/s_3)$, $\ud\phi_{12}=\ud\phi_1\ud\phi_2$, $\theta_{ij}=\phi_i-\phi_j$, $\Theta_{ij}^{e,\mu}=\theta_{ij}M_{ij}^{e,\mu2}$, where $M^e$ and $M^{\mu}$ are the effective mass~\cite{Kibble:1975vz} for the electron and muon, respectively,
\be
M_{ij}^{e2}=1+\langle{\bf a}^2\rangle_{ij}-\langle{\bf a}\rangle_{ij}^2
\qquad
M_{ij}^{\mu2}=\mu^2+\langle{\bf a}^2\rangle_{ij}-\langle{\bf a}\rangle_{ij}^2 \;,
\ee
where 
\be
\langle a\rangle_{ij}=\frac{1}{\theta_{ij}}\int_{\phi_i}^{\phi_j}\!\ud\phi\, a \;.
\ee
We have inserted factors of $s_0=1$ in appropriate places to make symmetries clearer.
As in other processes we have considered, we always find that the exponential part of the integrands can be expressed in terms of the effective mass.
The cross term between the ``lightfront-instantaneous'' and the ``three-point-vertex'' parts of the amplitude is give by
\be\label{P12dirGenFin}
\begin{split}
	\mathbb{P}_{12}(s)=\text{Re }i\frac{\alpha^2}{2\pi^2b_0}\int\!&\ud\phi_{123}\theta(\theta_{31})e^{\frac{i}{2b_0}\left[r_1\Theta_{21}^{\mu}+r_2\Theta_{23}^{e}\right]} \\
	&
	\frac{(s_0+s_1)(s_2-s_3)D_{12}}{q_1^3(\theta_{21}+i\epsilon)(\theta_{23}+i\epsilon)} \;,
\end{split}
\ee
where $D_{12}={\bf \Delta}_{12}\!\cdot\!{\bf \Delta}_{32}$ and 
\be
{\bf\Delta}_{ij}={\bf a}(\phi_i)-\langle{\bf a}\rangle_{ij} \;.
\ee
The third and final term is given by
\be\label{P22dirGenFin}
\begin{split}
	\mathbb{P}_{22}(s)=-&\frac{\alpha^2}{4\pi^2b_0^2}\int\!\ud^4\phi\frac{\theta(\theta_{31})\theta(\theta_{42})}{q_1^2\theta_{21}\theta_{43}} 
	e^{\frac{i}{2b_0}\left[r_1\Theta_{21}^{\mu}+r_2\Theta_{43}^{e}\right]} \\
	&\bigg\{\frac{\kappa_{01}\kappa_{23}}{4}W_{12}W_{34}+W_{13}W_{24}+W_{14}W_{23}
	\\
	&+\left[\frac{\kappa_{01}}{2}\left(\frac{2ib_0}{r_1\theta_{21}}+\mu^2+D_1\right)-\mu^2\right] \\
	&\left[\frac{\kappa_{23}}{2}\left(\frac{2ib_0}{r_2\theta_{43}}+1+D_2\right)+1\right]-D_1D_2\bigg\} \;,
\end{split}
\ee
where $\kappa_{ij}=(s_i/s_j)+(s_j/s_i)$, $D_1={\bf\Delta}_{12}\!\cdot\!{\bf\Delta}_{21}$, $D_2={\bf\Delta}_{34}\!\cdot\!{\bf\Delta}_{43}$, 
$W_{ij}=w_{i1}w_{j2}-w_{i2}w_{j1}=\hat{\bf z}\cdot({\bf w}_i\!\times\!{\bf w}_j)$,
${\bf w}_1={\bf\Delta}_{12}$, ${\bf w}_2={\bf\Delta}_{21}$, ${\bf w}_3={\bf\Delta}_{34}$ and 
${\bf w}_4={\bf\Delta}_{43}$.
In~\eqref{P22dirGenFin} and in the following we have left the $i\epsilon$ prescription implicit. The singularities at $\theta_{ij}=0$ are always avoided with an integration contour equivalent to replacing $\phi_{2,4}\to\phi_{2,4}+i\epsilon/2$ and $\phi_{1,3}\to\phi_{1,3}-i\epsilon/2$ with $\epsilon>0$. 

Note that $\mathbb{P}_{12}(s)$ is anti-symmetric with respect to $s_2\leftrightarrow s_3$, so for the integrated probability we find $\mathbb{P}_{12}=0$. Thus, for the integrated probability and for $\mu\ne1$, we just have two terms, $\mathbb{P}_{11}$ and $\mathbb{P}_{22}$. $\mathbb{P}_{11}$ is almost as simple as a first-order process, and $\mathbb{P}_{22}$ can be obtained from the incoherent product of the two first-order processes nonlinear Compton scattering by a ``muon'' and nonlinear Breit-Wheeler electron-positron pair production.
In some regimes, thought, it is natural to split $\mathbb{P}_{22}=\mathbb{P}_{22\to1}+\mathbb{P}_{22\to2}$ into two terms, where $\mathbb{P}_{22\to1}$ and $\mathbb{P}_{22\to2}$ scale linearly and quadratically in the volume, respectively. This can be done by splitting the step function combination in~\eqref{P22dirGenFin} as~\cite{Dinu:2017uoj}  
\be\label{StepsForStepsSep}
\begin{split}
&\theta(\theta_{42})\theta(\theta_{31}) \\
&=\theta(\sigma_{43}-\sigma_{21})\left\{1-\theta\left(\frac{|\theta_{43}-\theta_{21}|}{2}-[\sigma_{43}-\sigma_{21}]\right)\right\} \;.
\end{split}
\ee  
In the first term the average lightfront time in the pair-production step $\sigma_{43}=(\phi_4+\phi_3)/2$ can be much later than $\sigma_{21}=(\phi_2+\phi_1)/2$ for the photon-emission step, e.g. the photon can be emitted at one field maximum and then propagate to some later field maximum before it decays. In the second term $\sigma_{43}$ and $\sigma_{21}$ are forced to be close. So, the first term gives $\mathbb{P}_{22\to2}$ and the second term $\mathbb{P}_{22\to1}$. One can show that for $a_0\gg1$ we have $\mathbb{P}_{22\to2}=\mathcal{O}(a_0^2)+\mathcal{O}(a_0^0)$, while $\mathbb{P}_{22\to1}=\mathcal{O}(a_0)$, so this is a natural separation at least for large $a_0$. We therefore define 
\be
\mathbb{P}_{\rm one}=\mathbb{P}_{11}+\mathbb{P}_{22\to1} \qquad
\mathbb{P}_{\rm two}=\mathbb{P}_{22\to2} \;.
\ee  
The two-step $\mathbb{P}_{\rm two}$ gives the dominant contribution for high-intensity $a_0\gg1$ or for a long pulse length. This two-step dominance at $a_0\gg1$ is the basic assumption in particle-in-cell codes.  
In this paper we are interested in the high-energy limit, where the dominant contribution instead comes from the one-step $\mathbb{P}_{\rm one}$.


\section{High energy limit}\label{high-energy-section}

In this section we will study the limit where $b_0$ is the largest parameter in the system. In~\cite{Dinu:2019wdw} we showed numerically that the direct part of the one-step becomes dominant in this regime. In this section we will derive analytical approximations for this case. In the following two subsections we will for simplicity set $\mu=1$. We will reinstate $\mu$ in Sec.~\ref{CoulombSection}. 

\subsection{High-$\chi$ limit of LCF}\label{LCFsection}

For comparison, we first consider the large-$\chi$ limit of the familiar LCF. LCF can be obtained by starting with our expressions that are valid for arbitrary field shapes, and then expanding them in a power series in $1/a_0$, which is small in the LCF regime, see~\cite{Dinu:2017uoj}. We have $\mathbb{P}_{\rm two}=a_0^2P_2+\mathcal{O}(a_0^0)$ and $\mathbb{P}_{\rm one}\approx a_0P_1$. $P_2$ and $P_1$ depend on $b_0$ only via $\chi=a_0b_0$. For large $\chi$ we can neglect the exchange part of $P_1$ and we find
\be\label{PoneLCFchi}
\mathbb{P}_{\rm one}\approx\frac{13\alpha^2}{18\sqrt{3}\pi}\int\frac{\ud\sigma}{b_0}\chi(\sigma)\left(\ln\frac{\chi(\sigma)}{2\sqrt{3}}-\gamma_{\rm E}-\frac{142}{39}\right)
\ee 
and
\be\label{PtwoLCFchi}
\begin{split}
	\mathbb{P}_{\rm two}\approx&\frac{81\sqrt{3}}{56\pi^3}\Gamma^5\!\left[\frac{2}{3}\right]\frac{\alpha^2}{3^\frac{2}{3}}\int\!\frac{\ud\sigma_{21}}{b_0}\!\int_{\sigma_{21}}^\infty\!\frac{\ud\sigma_{43}}{b_0}(\chi(\sigma_{21})\chi(\sigma_{43}))^\frac{2}{3} \\
	&\times\left(\ln\chi(\sigma_{43})+\frac{7}{4}\ln3+\frac{7\pi}{4\sqrt{3}}-\gamma_{\rm E}-\frac{295}{42}\right) \;,
\end{split}
\ee
where $\gamma_{\rm E}\approx0.577$, $\chi(\sigma)=a_0b_0|f'(\sigma)|$ and the potential is given by $a(\sigma)=a_0f(\sigma)$. These are simple generalizations of the constant-crossed-field case, which was derived in~\cite{Ritus:1972nf}, to slowly varying, locally-constant fields. If we keep $a_0$ constant and increase $b_0$ then
$\mathbb{P}_{\rm one}\sim a_0\ln\chi$ and $\mathbb{P}_{\rm two}\sim a_0^2(1/\chi^{2/3})\ln\chi$ or 
\be\label{LCFratioHighchi}
\frac{\mathbb{P}_{\rm one}}{\mathbb{P}_{\rm two}}\sim\frac{\chi^{2/3}}{a_0} \;, 
\ee
which means that eventually (the right-hand-side of)~\eqref{PoneLCFchi} becomes larger than (the right-hand-side of)~\eqref{PtwoLCFchi}, suggesting that $\mathbb{P}_{\rm one}$ becomes larger than $\mathbb{P}_{\rm two}$ at sufficiently high energies. However, the LCF approximation breaks down at very high energies: One can obtain the LCF expansion by rescaling $\theta_{21}\to\tilde{\theta}_{21}/a_0$ and $\theta_{43}\to\tilde{\theta}_{43}/a_0$, and $\varphi=\sigma_{43}-\sigma_{21}\to\tilde{\varphi}/a_0$ for $\mathbb{P}_{\rm one}$, and then expanding the resulting integrands in $1/a_0$. In deriving the high-$\chi$ limit~\eqref{PoneLCFchi} and~\eqref{PtwoLCFchi} one finds that only a small fraction of the initial longitudinal momentum is given to the electron-positron pair, more precisely $s_{2,3}\sim1/\chi$. 
(Contrast this with the large $\chi$ limit of single-photon emission without subsequent pair production, where there instead is a peak where the emitted photon takes almost all the energy from the electron~\cite{Tamburini:2019tzo,Bulanov:2013cga}.)
One also finds that $\tilde{\theta}_{21}\sim\chi^{2/3}$. So, to be sure that the LCF expansion is still valid, we need 
\be\label{LCFconditionLargechi}
\frac{\chi^{2/3}}{a_0}\ll1  \;,
\ee
which means that for a given $a_0$ one cannot take $\chi$ arbitrarily large.
Note that this implies 
\be
a_0\gg b_0^2\gg1 \;,
\ee 
which is a more precise, process- and regime-specific condition compared to the general rule-of-thumb $a_0^2\gg b_0$~\cite{Dinu:2015aci}. 

As an aside we note that at low $b_0$ we find in general
\be
b_0\ll1: \qquad \mathbb{P}\sim\exp\left\{-\frac{f(a_0)}{\chi}\right\} \;,
\ee 
where $f(a_0)$ depends on the field shape. For $a_0\gg1$ we have
\be
b_0\ll1 \quad a_0\gg1: \qquad \mathbb{P}\sim\exp\left\{-\frac{16}{3\chi}+\frac{c}{\chi a_0^2}\right\} \;,
\ee
where the factor of $16/3\chi$ gives the LCF result and $c$ is a numerical factor. So, for LCF to be good in this regime one needs\footnote{For some field shapes, for example a circularly polarized monochromatic field, it happens that the numerical factor $c$ is actually small, which leads to a weaker condition so that one can use the LCF result even when $a_0\sim1$~\cite{NikishovRitusPairLinCirc,Hartin:2018sha,HernandezAcosta:2020agu}.}
\be
a_0^3\gg\frac{1}{b_0}\gg1 \;.
\ee

The condition~\eqref{LCFconditionLargechi} means that one cannot trust~\eqref{LCFratioHighchi} when this ratio becomes larger than one.
This is after all what one can expect for the relation between the first ($\mathbb{P}_{\rm two}$) and second term ($\mathbb{P}_{\rm one}$) in a (Laurent) power series.  
However, although the LCF approximation breaks down, we have found that if one increases $b_0$ with $a_0$ kept constant, then $\mathbb{P}_{\rm one}$ does in fact become larger than $\mathbb{P}_{\rm two}$ (but $\mathbb{P}_{\rm one}$ is no longer given by~\eqref{PoneLCFchi}). That $\mathbb{P}_{\rm one}$ can dominate in certain parameter regimes is not surprising, because for $a_0\ll1$ we have $\mathbb{P}_{\rm one}\sim\mathcal{O}(a_0^2)$ but $\mathbb{P}_{\rm two}\sim\mathcal{O}(a_0^4)$, so $\mathbb{P}_{\rm one}$ dominates for sufficiently weak fields. Moreover, the large-$b_0$ limit of the probabilities of the $\mathcal{O}(\alpha)$ processes nonlinear Compton and Breit-Wheeler~\cite{Podszus:2018hnz,Ilderton:2019kqp}, suggests that, at constant $a_0$ (but not necessarily small), increasing $b_0$ takes us closer to perturbative physics. In the following we will show that this is partly true for trident, but, due to the low momentum transfer, trident has a more nontrivial dependence on $a_0$.

\subsection{Large $b_0$ limit}\label{highEnergySection}

\begin{figure}
\includegraphics[width=\linewidth]{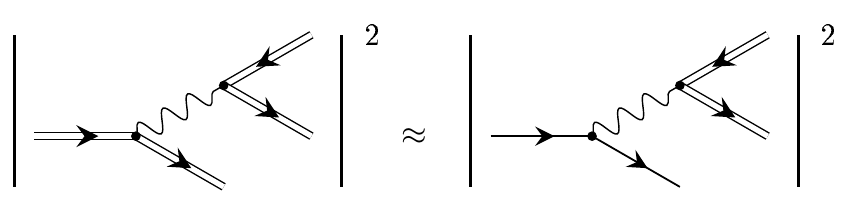}
\caption{Illustration of the fact that in the high-energy limit all the field dependence comes from the pair-production step. Double lines represent fermions dressed by the background field and the single line is an electron without interaction with the field. The wiggly line is a photon that is not part of the background field.}
\label{highEnergyTridentFig}
\end{figure}

So, we now take $b_0$ to be the largest parameter, and we therefore leave the LCF regime.
In this limit it turns out that the spectrum is peaked at $1-s_1\sim1/b_0\ll1$. This means that the initial electron keeps most of its longitudinal momentum, and the intermediate photon (in the $\mathbb{P}^{22\to1}$ case) scales as $kl=b_0q_1\sim\mathcal{O}(b_0^0)$, i.e.
\be
\frac{\chi_\gamma}{\chi}=\frac{kl}{kp}=\mathcal{O}(1/b_0) \;.
\ee
In the exponent of the direct terms we have factors of $r_1/b_0\sim1/b_0^2$, which suggests a rescaling of $\theta\to b_0^2\theta$ similar to~\cite{Ilderton:2019kqp} for single nonlinear Compton scattering. However, for the second step we have $r_2/b_0\sim b_0^0$, which leads to a fundamental difference. For $\mathbb{P}_{\rm dir}^{11}$ we should not rescale $\theta$ with $b_0$. In the limit of large $b_0$ we can still perform the longitudinal momentum integrals. We first change variables from $s_2=q_1(1+\nu)/2$ to $\nu$ and from $s_1=1/(1+t)$ to $t$. Then we rescale $t\to t/b_0$, expand the integrand in $b_0$ and perform the resulting $t$ and $\nu$ integrals. We find
\be\label{P11dirLargeb0}
\mathbb{P}_{\rm dir}^{11}=\frac{\alpha^2}{3\pi^2}\int\frac{\ud\sigma\ud\theta}{\theta\Theta}\frac{\partial\Theta}{\partial\theta} \;,
\ee  
with an integration contour equivalent to $\theta\to\theta+i\epsilon$. 
The second leading-order term comes from $\mathbb{P}_{\rm dir}^{22\to1}$. To calculate this term we start by making a partial integration in $\theta_{21}$ to change $1/\theta_{21}^2$ into $1/\theta_{21}$. In the non-boundary terms we rescale $\theta_{21}\to b_0^2\theta_{21}$ and $\sigma_{21}-\sigma_{43}\to b_0^2\sigma_{21}$. To leading order this means that there is no field dependence for the $\theta_{21}$ and $\sigma_{21}$ integrals, i.e. we can put e.g. $D_1\to0$. 
The $\sigma_{21}$ integral is trivial and gives a factor of $|\theta_{21}|$, which we represent as
\be
-\frac{|\theta_{21}|}{2}=\int\frac{\ud r}{\pi r^2}\left(e^{-\frac{i\theta_{21}r}{2}}-1\right) \;.
\ee     
Since $\Theta_{21}\to b_0^2\theta_{21}$ to leading order, the resulting $\theta_{21}$ integral can now be performed with the residue theorem. Then we perform the $r$ integral and finally the longitudinal momentum integrals ($t$ and $\nu$). The boundary term coming from the partial integration in $\theta_{21}$ is nonzero. To calculate this term we change variable from $\sigma_{21}$ to $\varphi=\sigma_{43}-\sigma_{21}$, write $1/(\varphi\pm\frac{\theta_{43}}{2})=\frac{\ud}{\ud\varphi}\ln(\varphi\pm\frac{\theta_{43}}{2})$ and make a partial integration in $\varphi$. The boundary term and the new $\varphi$ integral (in which we rescale $\varphi\to b_0^2\varphi$) can now be expanded in $b_0$. The longitudinal momentum integrals are again elementary. 
We thus find
\be\label{P221general}
\begin{split}
	\mathbb{P}_{\rm dir}^{22\to1}&=\frac{\alpha^2}{6\pi^2}\int\frac{\ud\sigma\ud\theta}{\theta\Theta}\bigg(9+\frac{19}{3}D_2-\frac{25}{3}\frac{\partial\Theta}{\partial\theta} \\
	&+[1-({\bf a}(\phi_3)-{\bf a}(\phi_4)^2)]\left[\ln\left[-\frac{\theta\Theta}{(2b_0)^2}\right]+2\gamma_{\rm E}\right]\bigg) \;,
\end{split}
\ee
where $\gamma_{\rm E}=0.577...$ is the Euler constant. We have used $({\bf a}(\phi_3)-{\bf a}(\phi_4)^2=2(\partial_\theta\Theta-D_2-1)$. Using this relation again we find that the total probability $\mathbb{P}\approx\mathbb{P}_{\rm dir}^{11}+\mathbb{P}_{\rm dir}^{22\to1}$ is given by
\be\label{totalPgeneral}
\begin{split}
	\mathbb{P}&=\frac{\alpha^2}{6\pi^2}\int\frac{\ud\phi_3\ud\phi_4}{\theta\Theta}\bigg(\frac{8}{3}-\frac{19}{6}({\bf a}(\phi_4)-{\bf a}(\phi_3))^2 \\
	&+[1-({\bf a}(\phi_4)-{\bf a}(\phi_3))^2]\left[\ln\left[-\frac{\theta\Theta}{(2b_0)^2}\right]+2\gamma_{\rm E}\right]\bigg) \;.
\end{split}
\ee
Note that, unlike the probabilities for the first-order processes nonlinear Compton and Breit-Wheeler~\cite{Podszus:2018hnz,Ilderton:2019kqp}, $\mathbb{P}$ does not scale as $a_0^2$; it has instead a nontrivial dependence on $a_0$\footnote{Nontrivial dependences on $a_0$ in the high-energy limit have also been found in the real parts of the two loop diagrams, whose imaginary parts give the probabilities of nonlinear Compton and Breit-Wheeler~\cite{Podszus:2018hnz}.}.
The reason for this is that, while the incoming particle in the first step has high energy (which leads to perturbative scalings), the particles involved in the second step do not. This suggests that higher-order diagrams will in general have subprocesses associated with lower energy which give more nontrivial dependencies on the field strength. The appearance of softer ($\chi\sim1$) vertices is of course also what one would expect for late vertices in cascades, after the initial momentum has been distributed among a large number of particles. However, here we see that this happens already at $\mathcal{O}(\alpha^2)$.  

Although~\eqref{totalPgeneral} has in general a nonlinear dependence on $a_0$, the first step in the trident process is simple in a way similar to the first-order probabilities in~\cite{Podszus:2018hnz,Ilderton:2019kqp}. In fact, the first step in trident has to leading order no dependence on the field, all the field dependence comes from the second step, see Fig.~\ref{highEnergyTridentFig}. This generalizes a corresponding result in perturbative $\mathcal{O}(a_0^2)$ trident~\cite{Borsellino,Kopylov63}, where a single photon (which would come from the background field in our case) is absorbed, and where to leading order this photon is absorbed only by the pair-production step. For small $a_0$ we can expand~\eqref{totalPgeneral} to $\mathcal{O}(a_0^2)$ and compare with the old literature on perturbative trident. We find
\be\label{perturbativeLargeb0}
\mathbb{P}=\frac{\alpha^2}{2\pi}\int_0^\infty\frac{\ud w}{2\pi}|a(w)|^2w\left(\frac{28}{9}\ln(2b_0w)-\frac{218}{27}\right) \;,
\ee 
where the Fourier transform of the field is given by
\be
a(w)=\int\!\ud\phi\; a(\phi)e^{iw\phi} \;.
\ee   
($\mathbb{P}_{\rm dir}^{11}$ contributes a factor of $-6/109\approx-0.06$ of the non-log term in~\eqref{perturbativeLargeb0}.) To compare the probability~\eqref{perturbativeLargeb0} with the cross section in the literature, we replace the Fourier transform $a(w)\to e\epsilon_\mu2\pi\delta(w-w_0)/\sqrt{2\omega V_3}$ and divide by the flux density ($1/V_3$) and a temporal volume factor. We then recover exactly the literature result, see~\cite{Borsellino,Kopylov63}.    

To compare with the LCF result~\eqref{PoneLCFchi}, let us consider the limit $a_0\gg1$. One should not expect the $a_0\gg1$ limit of our high-$b_0$ approximation~\eqref{totalPgeneral} to reduce to the large-$\chi$ limit of LCF~\eqref{PoneLCFchi}, because taking $b_0$ to be largest and then $a_0$ to be large does not commute with taking $a_0$ to be largest and then $b_0$ to be large~\cite{Podszus:2018hnz,Ilderton:2019kqp}. We find by taking the $a_0\gg1$ limit of~\eqref{P221general} and~\eqref{P11dirLargeb0}
\be\label{LCFlimitP11}
\mathbb{P}_{\rm dir}^{11}=\frac{\alpha^2}{3\sqrt{3}\pi}\int\frac{\ud\sigma}{b_0}\chi(\sigma)
\ee   
and
\be
\mathbb{P}_{\rm dir}^{22\to1}=\frac{13\alpha^2}{6\sqrt{3}\pi}\int\frac{\ud\sigma}{b_0}\chi(\sigma)\left(\ln\frac{\chi(\sigma)}{2\sqrt{3}}-\gamma_{\rm E}-\frac{64}{39}\right) \;.
\ee
The total probability is thus given by
\be\label{PLCF}
\mathbb{P}=\frac{13\alpha^2}{6\sqrt{3}\pi}\int\frac{\ud\sigma}{b_0}\chi(\sigma)\left(\ln\frac{\chi(\sigma)}{2\sqrt{3}}-\gamma_{\rm E}-\frac{58}{39}\right) \;.
\ee
Although this is different from the $\chi\gg1$ limit of the LCF approximation~\eqref{PoneLCFchi}, it nevertheless looks quite similar. There is, however, an important difference.

\subsection{Muon trident and pair production by a Coulomb field}\label{CoulombSection}

This difference is not obvious in the above expressions, but it becomes obvious if we replace the initial electron with a muon (or some other lepton with a different mass) with mass $\mu\ne1$ (we still use units where the electron mass $m_e=1$).
The LCF approximation~\eqref{PoneLCFchi} is independent of $\mu$, see~\cite{Ritus:1972nf}.
In contrast, the generalization of~\eqref{totalPgeneral} is obtained in the same way as in the perturbative case, i.e. one should replace $b_0\to b_0/\mu$. This means that in the rest frame of the initial particle, the probability is independent of $\mu$. This suggests that~\eqref{PLCF} can be directly compared with the probability of pair production by a plane-wave field and an infinitely massive initial particle in the form of a stationary Coulomb field. That process has been calculated in a constant-crossed field in~\cite{CoulombNarozhnyiNikishov,Ritus:1972nf}. We find that~\eqref{PLCF} agrees perfectly with Eq.~(19) in~\cite{CoulombNarozhnyiNikishov} or with Eq.~(45) in~\cite{Ritus:1972nf} ($2\text{Im} T$ gives the pair-production probability). 
Thus, our new large-$b_0$ approximation interpolates between the old result for perturbative trident for $a_0\ll1$ and the old result for pair production by a Coulomb field in a constant-crossed field for $a_0\gg1$. 
This relation with pair production by a Coulomb field can also be seen in the perturbative case $\mathcal{O}(a_0^2)$ \cite{Borsellino,SuhBethe,JauchRohrlich}, so one can expect it to hold for arbitrary $a_0$. 

To show that this is indeed the case, we need to calculate the probability of pair production by a Coulomb field and an inhomogeneous plane wave with arbitrary $a_0$.
Formally, the calculation is similar to nonlinear Breit-Wheeler pair production, except that the photon polarization vector $\epsilon_\mu$ should be replaced by the Fourier transform of the Coulomb field $\mathcal{A}_0(x)=\frac{e}{4\pi r}$, which is given by $\mathcal{A}_0(l)=e/{\bf l}^2$, and the Coulomb photon is off shell. The amplitude is given by
\be
\mathcal{M}=-ie\int\frac{\ud^3{\bf l}}{(2\pi)^3}\int\ud^4x\,\underset{p_2}{\bar{\psi}(x)}\mathcal{A}({\bf l})\gamma^0e^{-il_ix^i}\underset{p_3}{\psi_\LCm(x)} \;,
\ee   
where $\psi$ and $\psi_\LCm$ are the Volkov solutions for the electron and positron (same notation as in~\cite{Dinu:2017uoj}). We work in the rest frame of the initial particle (the Coulomb center), so $b_0$ gives the frequency of the plane wave.
The integrals over $x^\LCperp$ and $x^\LCm$ give a delta function which we use to perform the integrals over $l_\LCperp$ and $l_3=-2l_\LCm=2l_\LCp$. 
The probability is given by
\be
\mathbb{P}=\int\ud\tilde{p}_2\ud\tilde{p}_3|M|^2 \;,
\ee
where $\ud\tilde{p}=\theta(p_\LCm)\ud p_\LCm\ud^2p_\LCperp/(2p_\LCm(2\pi)^3)$ is Lorentz invariant, and $p_2$ and $p_3$ are the momenta of the electron and positron.
We exponentiate the Coulomb factor
\be
\frac{1}{{\bf l}^4}=-\left(\frac{\theta}{2b_0}\right)^2\int\ud u\, u\exp\left(\frac{i\theta}{2b_0}{\bf l}^2u\right)
\ee
and perform the resulting Gaussian integrals over $p_{2\LCperp}$ and $p_{3\LCperp}$. We are using the Coulomb gauge for the Coulomb field, rather than the lightfront gauge, and we find terms that are conveniently rewritten using partial integration, using e.g. 
\be
\frac{\partial\Theta}{\partial\phi_3}=-(1+{\bf w}_3^2)
\qquad
\frac{\partial\Theta}{\partial\phi_4}=1+{\bf w}_4^2 \;.
\ee 
Next we can perform the $u$ integral in terms of an incomplete gamma function,
\be\label{PExactCoulomb}
\begin{split}
\mathbb{P}(s)&=\frac{\alpha^2}{4\pi^2b_0^2q_1^2}\int\ud\phi_3\ud\phi_4\exp\left\{\frac{iq_1\theta}{2b_0}+\frac{ir\Theta}{2b_0}\right\} \\
&\bigg\{-\frac{8s_2s_3b_0^2}{q_1^2\theta^2}+ \\
&\left[\frac{2ib_0}{q_1\theta}(\mathcal{J}-1)+\mathcal{J}\right]\left[\frac{\kappa}{2}\left(\frac{2ib_0}{r\theta}+D_2+1\right)+1\right]\bigg\} \;,
\end{split}
\ee
where $r=(1/s_2)+(1/s_3)$, $\kappa=(s_2/s_3)+(s_3/s_2)$, $q_1=s_2+s_3$ and
\be
\mathcal{J}=e^{iz}\Gamma(0,iz) \qquad z=-\frac{(s_2+s_3)\theta}{2b_0} \;.
\ee
This result~\eqref{PExactCoulomb} is exact in $b_0$ (and $a_0$).
For large $b_0$ it gives logarithmic terms. The large $b_0$ limit is obtained by rescaling $s_{2,3}\to s_{2,3}/b_0$ and expanding in $b_0$. We change variables from $s_2=t(1-\nu)/2$ and $s_3=t(1+\nu)/2$ to $t$ and $\nu$. After performing these (elementary) longitudinal momentum integrals we finally find~\eqref{totalPgeneral}. Thus, our high-energy approximation~\eqref{totalPgeneral} for trident agrees exactly with the high-energy limit of pair production by a Coulomb $+$ plane-wave field, for arbitrary $a_0$, field shape and polarization. Although in the near future it will probably be easier to reach high $\chi$ by increasing $a_0$ rather than $b_0$, i.e. within the LCF regime, the high-$\chi$ limit of LCF does not agree with the result for Coulomb $+$ constant-crossed field, this connection is instead seen in the high-$b_0$ limit~\eqref{totalPgeneral}.

Pair production by the combination of a Coulomb field and inhomogeneous plane waves has been studied at high energies in~\cite{Yakovlev,Milstein:2006zz,DiPiazza:2009py,DiPiazza:2009yi,DiPiazza:2010kg}. For a comparison between muon trident and pair production by a Coulomb field in a plane wave see~\cite{Muller:2009ri}.

\subsection{Weizs\"acker-Williams equivalent photon approximation}\label{WWsection}

For other processes, in the absence of a strong laser field, a common tool for studying the high-energy limit is the Weizs\"acker-Williams (WW) equivalent photon approximation~\cite{Weizsacker,Williams,FermiWW}, see e.g.~\cite{QED-book} for a textbook treatment.
At the time of the famous experiment at SLAC~\cite{Bamber:1999zt}, no complete description of trident existed, so a WW approximation was used to estimate the importance of the one-step term\footnote{The one-step term was called trident in~\cite{Bamber:1999zt}, but we use trident to refer to the total probability.}. However, in~\cite{King:2013osa} it was shown that the WW approach does not agree with the high-$\chi$ limit of the LCF approximations. In this section we will explain why this is.  
 
In our case the starting point for a WW approximation is given by (cf.~\cite{King:2013osa})
\be
\mathbb{P}_{\rm WW}=\frac{2\alpha}{\pi}\int\frac{\ud q_1}{q_1}\ln\left(\frac{1}{q_1}\right)\mathbb{P}_{\rm BW} \;,
\ee
where $\mathbb{P}_{\rm BW}$ is the photon-averaged probability of nonlinear Breit-Wheeler pair production, which can be expressed as~\cite{Dinu:2017uoj}
\be\label{BWfromPhi3Phi4}
\begin{split}
	\mathbb{P}_{\rm BW}=\frac{i\alpha}{2\pi b_0}\int_0^{q_1}\ud s_2&\int\frac{\ud\phi_3\ud\phi_4}{q_1^2\theta_{43}}\exp\left\{\frac{ir_2}{2b_0}\Theta_{43}\right\} \\
	&\left(1-\frac{\kappa_{23}}{4}({\bf a}(\phi_4)-{\bf a}(\phi_3))^2\right) \;,
\end{split}
\ee
where again $q_1=s_2+s_3$ is the longitudinal photon momentum.
We again rescale $s_{2,3}\to s_{2,3}/b_0$ and change variables from $s_2=t(1-\nu)/2$ and $s_3=t(1+\nu)/2$ to $t$ and $\nu$. To leading (logarithmic) order we find
\be
\mathbb{P}_{\rm WW}=\frac{\alpha^2}{3\pi^2}\int\frac{\ud\phi_3\phi_4}{\theta\Theta}[({\bf a}(\phi_4)-{\bf a}(\phi_3))^2-1]\ln b_0 \;,
\ee  
which is exactly the same as the $\ln b_0$ part of the full approximation~\eqref{totalPgeneral}. 
So, the WW approach does work. It agrees with our new approximation where $b_0$ is the largest parameter. Since the large $a_0$ limit of this approximation does not commute with the large $b_0$ limit of the LCF approximation, we now see that the reason that WW and LCF do not agree is that for WW to work we need $b_0$ to be the largest parameter, while for LCF to work we need $a_0$ to be largest.
The WW approach might be the simplest way to obtain the $\ln b_0$ term, but $b_0$ would have to be very large in order for the constant terms to be negligible compared to this logarithmic term.

\subsection{Nonlocal corrections}\label{nextToLead}

\begin{figure}
\includegraphics[width=\linewidth]{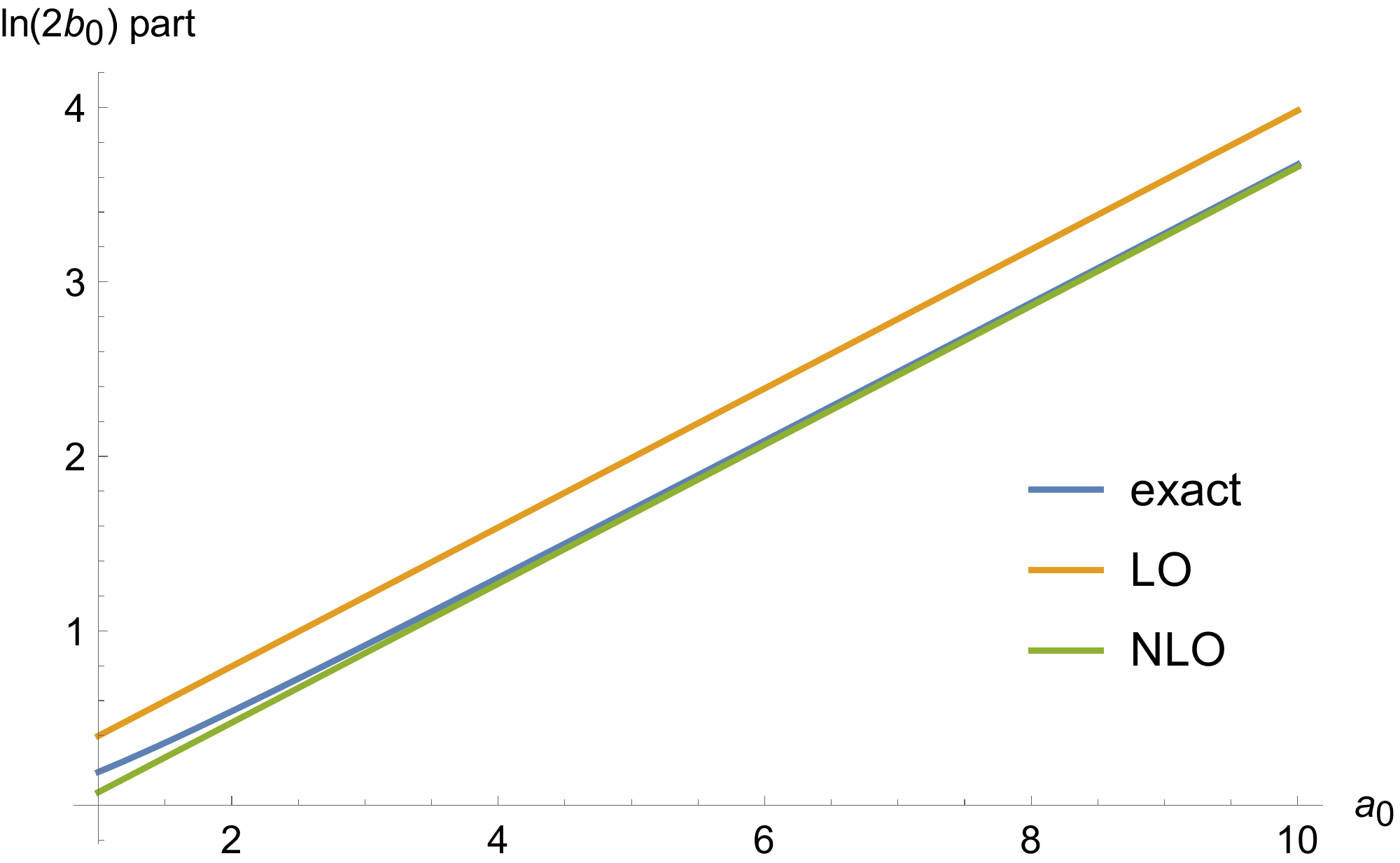}
\includegraphics[width=\linewidth]{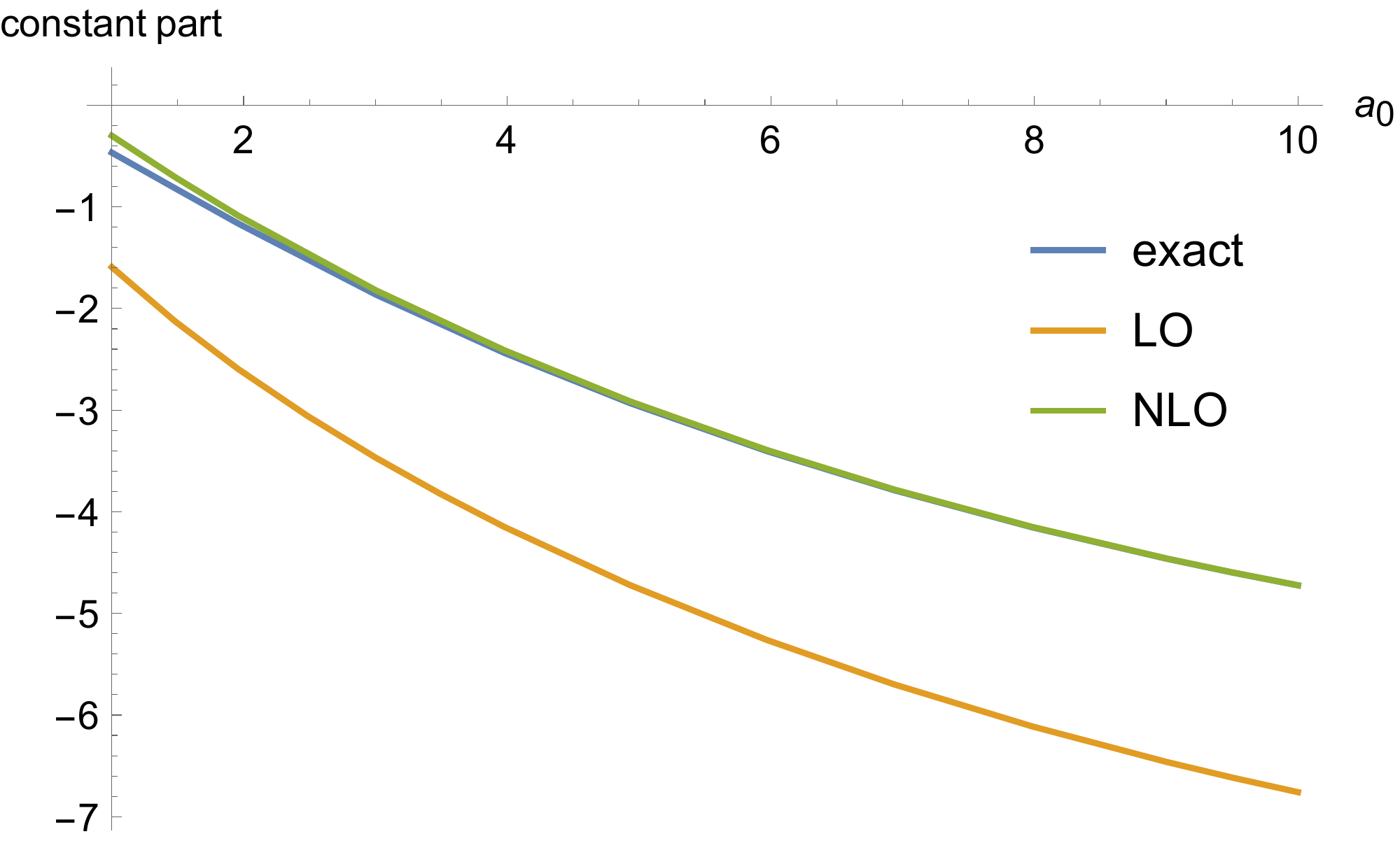}
\caption{Comparison between the exact, the leading order~\eqref{PLCF}, and the leading order plus the next-to-leading order correction~\eqref{NLOcirc}. The field is monochromatic with a circular polarization, ${\bf f}=\{\sin\phi,\cos\phi\}$. The first plot shows the part proportional to $\ln2b_0$ and the second plot the rest. In both cases a factor of $\alpha^2\mathcal{T}$ has been factored out, where $\mathcal{T}$ is a volume factor. The nonlocal correction is clearly important here.}
\label{NLOcirc}
\end{figure}

\begin{figure}
\includegraphics[width=\linewidth]{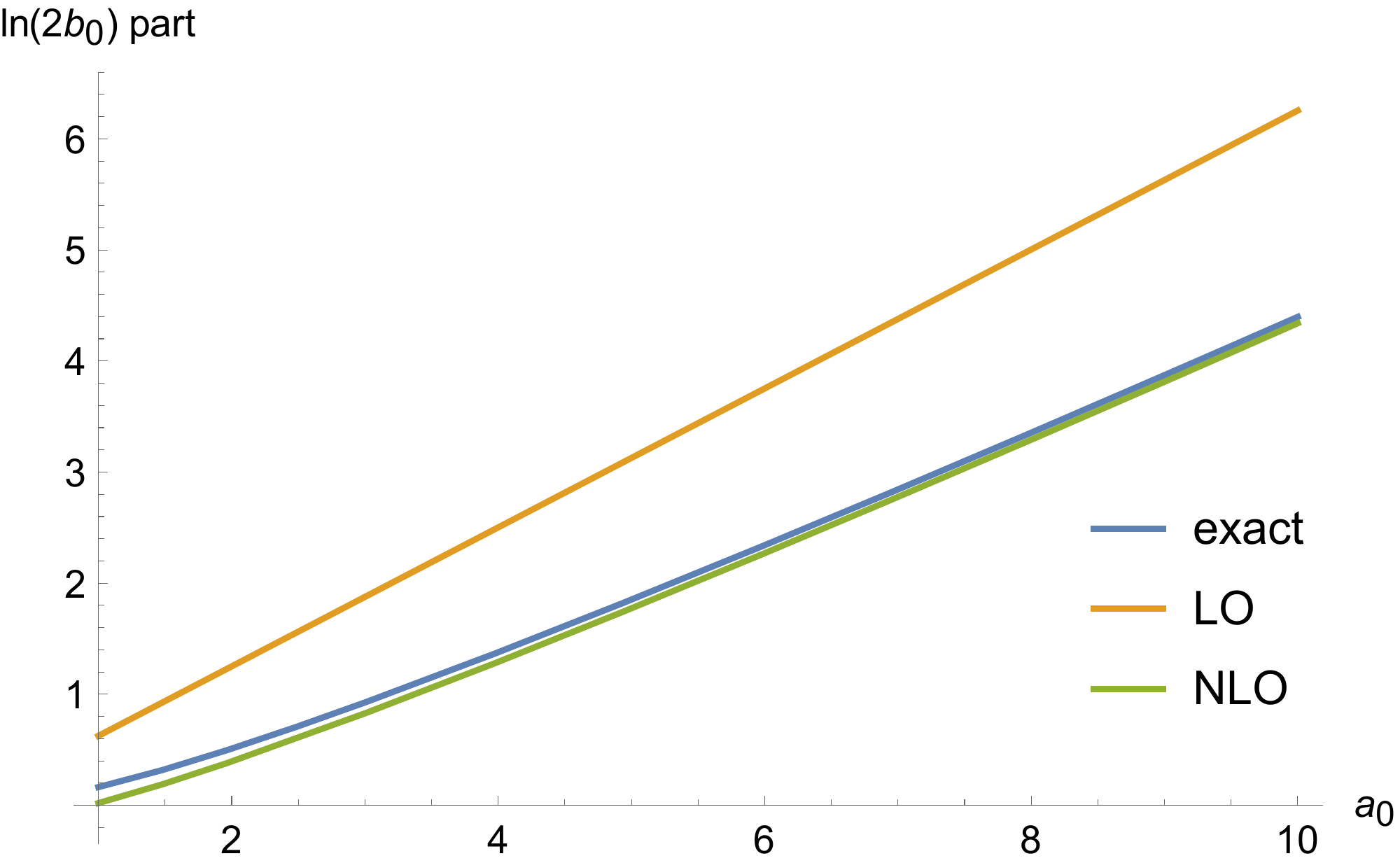}
\includegraphics[width=\linewidth]{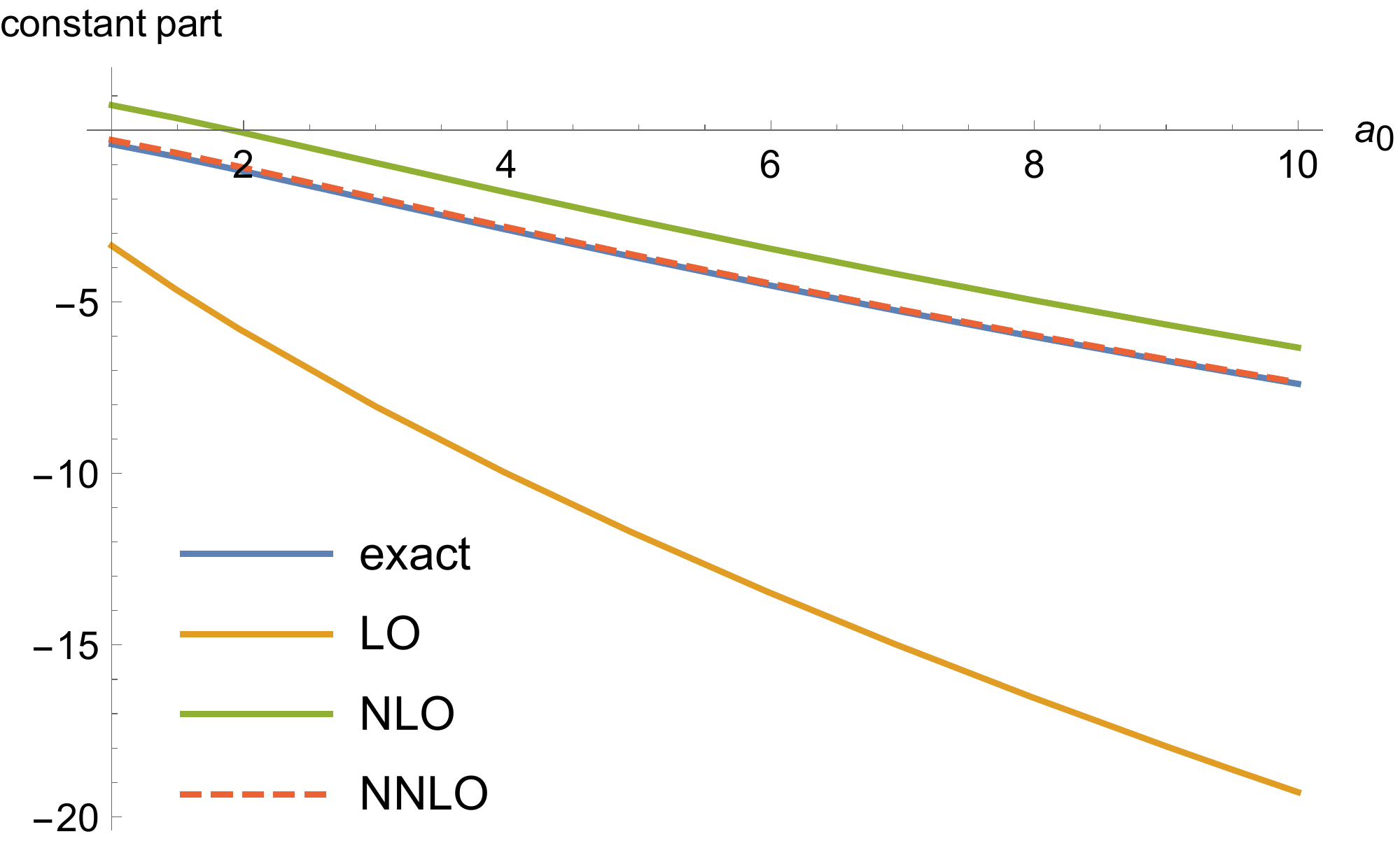}
\caption{Same as~\ref{NLOcirc} but with a Lorentzian ($n=1$ in~\eqref{LorentzDefinition}) instead of a flat-top envelope. Since LO$\sim a_0$ and NLO$\sim\sqrt{a_0}$ (apart from log terms), the absolute difference increases with $a_0$, so NLO is even more important for this pulse shape. For the constant part the NNLO scales as $a_0^0$.}
\label{NLOcircLor}
\end{figure}

In Sec.~\ref{CoulombSection} we showed that the large-$a_0$ limit of our large-$b_0$ approximation is fundamentally different from the large-$\chi$ limit of the LCF approximation, even though they at first sight look similar. 
This difference becomes even clearer at the next-to-leading order (NLO).  
The leading order (LO)~\eqref{PLCF} is obtained in a way that is similar to the derivation of LCF, i.e. it is obtained by rescaling $\theta\to\tilde{\theta}/a_0$ and then expanding to leading order in $1/a_0$. This is a local, derivative expansion around the point where $\theta=0$ or $\phi_3=\phi_4$. For example, the effective mass becomes $M^2\approx1+\dot{a}_\LCperp^2(\sigma)\theta^2/12$ with corrections involving higher derivatives of $a(\sigma)$.
The correction to the leading-order LCF approximation~\eqref{PoneLCFchi} and~\eqref{PtwoLCFchi} is obtained by simply including higher orders in this expansion. In contrast, we will now show that the next-to-leading order correction to~\eqref{PLCF} is nonlocal. In fact, (at least for a long pulse considered in this section) its scaling with respect to $a_0$ is not universal, it depends on the pulse shape. 

We can see this using a long pulse with circular polarization, $a(\phi)=a_0(\sin\phi,\cos\phi)h(\phi/\mathcal{T})$, where $h(x)$ gives the envelope shape, e.g. $e^{-x^2}$ or $\theta(1-2|x|)$. For $\mathcal{T}\gg1$ we rescale $\sigma=\mathcal{T}u$ and expand in $\mathcal{T}$. (The locally monochromatic approximation has recently been studied in~\cite{Heinzl:2020ynb}.) We have
\be
M^2\approx1+[a_0h(u)]^2\left[1-\text{sinc}^2\frac{\theta}{2}\right]
\ee
and
\be
({\bf a}(\phi_4)-{\bf a}(\phi_3))^2\approx[a_0h(u)]^2\theta^2\text{sinc}^2\frac{\theta}{2} \;.
\ee
We see that $u$ only appears in the integrand via $a_0h(u)$. Let us for simplicity consider first a flat-top envelope $h(x)=\theta(1-2|x|)$, so the $u$ integral gives trivially $1$.
We obtain NLO by subtracting from the exact integrand the integrand that gives LO (which is obtained by rescaling $\theta\to\tilde{\theta}/a_0$ and expanding to leading order in $1/a_0$), but expressed in terms of the original $\theta$ rather than $\tilde{\theta}=a_0\theta$, and then we expand this difference directly in $1/a_0$, i.e. without rescaling any integration variables. 
We find
\be\label{totalPnlo}
\Delta\mathbb{P}=-\frac{\alpha^2}{6\pi^2}\mathcal{T}\int\ud\theta(\mathcal{F}-\mathcal{F}_{\rm LCF}) \;,
\ee
where
\be
\mathcal{F}=\frac{({\bf f}(\phi_4)-{\bf f}(\phi_3))^2}{\theta^2F}
\left(\frac{19}{6}+\ln\left[\frac{a_0^2\theta^2F}{(2b_0)^2}\right]+2\gamma_{\rm E}\right) \;,
\ee
\be
\mathcal{F}_{\rm LCF}=\frac{12}{\theta^2}
\left(\frac{19}{6}+\ln\left[\frac{a_0^2\theta^2F_{\rm LCF}}{(2b_0)^2}\right]+2\gamma_{\rm E}\right) \;,
\ee
\be
F=\langle{\bf f}^2\rangle-\langle{\bf f}\rangle^2\approx1-\text{sinc}^2\frac{\theta}{2} \;,
\ee
\be
\frac{({\bf f}(\phi_4)-{\bf f}(\phi_3))^2}{\theta^2}\approx\text{sinc}^2\frac{\theta}{2} \;,
\ee
and $F_{\rm LCF}=\dot{{\bf f}}^2(\sigma)\theta^2/12$.
The integrand in~\eqref{totalPnlo} has an integrable singularity at $\theta=0$, so we can set $i\epsilon\to0$. Note that, in contrast to LO~\eqref{PLCF} and what one might have expected from the LCF regime, this NLO depends nonlocally on the field, i.e. it is not an expansion around $\theta=0$ and the dominant contribution to the integral comes from a $\theta$ interval with $\theta\sim1$ (neither large nor small, dimensionless).
So, while we for LO~\eqref{PLCF} can perform the $\theta$ integral for an arbitrary field shape, in NLO we still have a nontrivial $\theta$ integral that feels all of the field shape. 
Since we are in a regime where $b_0$ is supposed to be larger than any other parameter, one might have expected that the formation length should be large and then the nonlocality would not be surprising. But note that the dependence on $a_0$ and $b_0$ in~\eqref{totalPgeneral} is separated into $h(a_0)\ln b_0+g(a_0)$, where $h(a_0)$ and $g(a_0)$ only depend on $a_0$ and the pulse shape. So, whether or not we can approximate the functions $g(a_0)$ and $h(a_0)$ using a local $\theta\sim1/a_0$ scaling is not determined by $b_0$. And if $a_0$ is sufficiently large (but with $b_0$ still being large enough such that~\eqref{totalPgeneral} is valid) the leading order does still come from a short formation length $\theta\sim1/a_0$ independently of $b_0$.

Note also that NLO scales as $a_0^0$ and $\ln a_0$ compared LO which scales as $a_0$ and $a_0\ln a_0$, so, unless $a_0$ is very large, \eqref{totalPnlo} provides a numerically important correction. This is illustrated in Fig.~\ref{NLOcirc}. 
For this field we find
\be
\begin{split}
\mathbb{P}=&\alpha^2\mathcal{T}(0.40a_0-0.32)\ln(2b_0) \\
&+\alpha^2\mathcal{T}(-1.6a_0+1.3+[0.32+0.40a_0]\ln a_0) \;,
\end{split}
\ee
where all the numerical factors are approximate.
At least for this example, Fig.~\ref{NLOcirc} shows that by including NLO we have a good approximation already at $a_0>2$.


If we instead of a flat-top envelope $h(x)=\theta(1-2|x|)$ have a smooth envelope, then the $u$ integrand is approximately constant, equal to~\eqref{totalPnlo}, in the interval where $a_0h(u)\gg1$, but the fact that the length of this interval is now $a_0$ dependent means that NLO has a different scaling with respect to $a_0$. Consider for example 
\be\label{LorentzDefinition}
h(x)=1/(1+[2x]^{2n})
\ee
with $n\geq1$. In the limit $n\to\infty$ we recover the flat-top $h(x)=\theta(1-2|x|)$.
We obtain the NLO by subtracting the LO integrand, as in~\eqref{totalPnlo}, except this time we rescale $u\to a_0^{1/(2n)}\tilde{u}$ before we take the limit $a_0\gg1$. We find that NLO scales as $a_0^{1/(2n)}$ (with some terms having an additional $\ln a_0$). This means that for a smooth envelope NLO is even more important. It is most important for the field with the slowest decay, $n=1$, where the ratio between LO and NLO only scale as $\sqrt{a_0}$.
Note that LO is obtained from $\phi$ values on the order $\sigma\sim a_0^0$ and $\theta\sim1/a_0$, while NLO is obtained from $\sigma\sim a_0^{1/(2n)}$ and $\theta\sim a_0^0$. 
Note also that the scaling with respect to $a_0$ is not universal, it depends on the pulse shape, $n$ in this case. This also highlights the fact that NLO is not simply the next term in a power-series expansion in $1/a_0$ (in contrast to the LCF regime).

Fig.~\ref{NLOcircLor} shows that for a Lorentzian pulse ($n=1$) shape the NLO term is indeed more important than for the flat-top envelope. This is especially clear for the $b_0$-independent term, for which the error at leading order is even larger than the exact result, even for $a_0=10$.
For the $b_0$-independent term the NNLO term is a constant, $a_0^0$. This NNLO term is obtained in the following way. Let $\mathcal{I}(\sigma,\theta)$ be the integrand. LO is obtained by rescaling $\theta\to\tilde{\theta}/a_0$ and expanding to leading order in $a_0$. If one tried to obtain the next order by keeping $\sigma$ and $\tilde{\theta}$ as independent of $a_0$ then one would find divergent integrals, so NLO must instead be obtained by a different $a_0$ scaling of the integration variables. Let $\mathcal{I}_{\rm LO}(\sigma,\theta)$ be the leading-order integrand expressed in terms of the original $\theta$ variable. NLO is now obtained from $\mathcal{I}(\sigma,\theta)-\mathcal{I}_{\rm LO}(\sigma,\theta)$ by rescaling $\sigma\to\sqrt{a_0}\tilde{\sigma}$ and expanding to leading order in $a_0$. Again, one cannot obtain NNLO with the same $a_0$ scaling of $\theta$ and $\sigma$, because this leads to divergent integrals. Let $\mathcal{I}_{\rm NLO}(\sigma,\theta)$ be the integrand that gives NLO, expressed in terms of the original variables. NNLO is now obtained from $(\mathcal{I}-\mathcal{I}_{\rm LO}-\mathcal{I}_{\rm LO})(\sigma,\theta)$ by expanding to leading order in $a_0$, this time without rescaling the integration variables. So, each of these terms are obtained with a different rescaling of $\sigma$ and $\theta$. Contrast this with the LCF or the saddle-point regime, where one just have to rescale the integration variables once and then obtain the leading as well as higher orders by simply expanding the integrand to higher orders. In the saddle-point regime one would for example change variables to $\theta=\theta_{\rm saddle}+\sqrt{\chi}\tilde{\theta}$ etc. and then expand the integrand in a power series in $\chi$ times and exponential on the form $e^{-.../\chi}$. This leads in general to an asymptotic series, but the terms are obtained in a systematic way. Here we need to work more in order to obtain the higher orders in the large-$a_0$ expansion, and, moreover, the scalings of NLO and NNLO are not universal, they depend on the field shape.
We also see that, unless $a_0$ is very large, LO can be far from the exact result, which means that we need to obtain these higher orders. Fortunately, once we have calculated the first orders we obtain a very good approximation already at $a_0\gtrsim2$.

\subsection{Nonlocal corrections for short pulses}\label{nextToLeadShort}

In the previous section we showed that NLO can be important for a long pulse. However, the longer the pulse is, the higher the energy has to be for our high-energy approximation~\eqref{totalPgeneral} to good. So, in this section we will study NLO for short pulses. Since unipolar fields seem to involve more complicated calculations, we focus on a field with $a(-\infty)=a(\infty)$, given by 
\be\label{compactField}
a(\phi)=a_0(1+\cos\phi)
\ee
for $|\phi|<\pi$ and $a(\phi)=0$ for $|\phi|>\pi$. As in the previous section, LO is obtained by expanding to leading order in $\theta$, while the corrections are nonlocal. Due to symmetry we can restrict the integration variables to $\phi_1<\phi_2$. We separate the integration region into parts with $\mathcal{P}_{12}=\{\phi_1<-\pi,-\pi<\phi_2<\pi\}$, $\mathcal{P}_{22}=\{-\pi<\phi_1<\phi_2<\pi\}$, $\mathcal{P}_{23}=\{-\pi<\phi_1<\pi,\phi_2>\pi\}$ and $\mathcal{P}_{13}=\{\phi_1<-\pi,\phi_2>\pi\}$. The contribution from $\mathcal{P}_{23}$ is equal to the one from $\mathcal{P}_{12}$, and only $\mathcal{P}_{22}$ contribute to LO. NLO can be calculated in a similar way as in the previous section, but the fact that we have two nontrivial integrals makes the calculations more complicated. So, we simply state the result
\be\label{compactExample1}
\begin{split}
\mathbb{P}\approx\alpha^2&\Big[(1.6a_0+0.34[\ln a_0]^2+0.25\ln a_0-0.99)\ln(2b_0) \\ 
&+1.6a_0\ln a_0-6.9a_0 \\
&-0.28[\ln a_0]^3-1.6[\ln a_0]^2-0.32\ln a_0+4.9\Big] \;,
\end{split}
\ee   
where all the constants are approximate. While a direct calculation of this is quite involved, we can confirm it more easily by making the following ansatz for the correction,
\be\label{logAnsatz}
\delta\mathbb{P}(a_0)=d_0+d_1 \ln a_0+d_2\left[\ln a_0\right]^2+d_3\left[\ln a_0\right]^3 \;,
\ee
where $d_i$ are constants that can be obtained either by a numerical evaluation of the exact result for $\mathbb{P}$, $\ud\mathbb{P}/\ud a_0$, $\ud^2\mathbb{P}/\ud^2 a_0$ and $\ud^3\mathbb{P}/\ud^3 a_0$ at one, arbitrary, large $a_0=a_r$; or by evaluating $\mathbb{P}$ at 4 different $a_r$.
$a_r$ should be large enough so that the exact result has converged to~\eqref{logAnsatz}, and can be chosen much larger than the $a_0$ range one is mainly interested in. In this case we have checked that $a_r\sim10^4$ gives good results: 

The need for NLO is most clearly seen in the $b_0$-independent part. At $a_0=30$ the exact result for this part is $\approx-69.41\alpha^2$, which is in good agreement with LO$+$NLO$\approx-69.45\alpha^2$, while the leading order is not great, LO$\approx-43.27\alpha^2$. This is consistent with the results in the previous section for long pulses, i.e. that NLO is needed to have a good precision even for very large $a_0$. Moreover, we again find that by including NLO we have a good approximation already at not-very-large $a_0$; for $a_0=2$ we find for the $b_0$-independent part $\{\text{exact},\text{LO},\text{LO}+\text{NLO}\}\approx\{-7.9,-11.5,-7.7\}\alpha^2$, so LO$+$NLO is already close to the exact result at $a_0=2$. 

A short pulse with compact support is also useful in order to demonstrate that the correction is nonlocal, because the part where the two integration variables are both outside the pulse but on opposite sides, i.e. $\mathcal{P}_{13}$, contributes to NLO. In this example we have
\be
\begin{split}
\delta\mathbb{P}_{13}\approx&\alpha^2\Big[(0.14\ln a_0-0.040)\ln(2b_0) \\
&-0.14[\ln a_0]^2-0.56\ln a_0+0.16\Big] \;,
\end{split}
\ee
which is a significant part of the total NLO.

Given that NLO is nonlocal, one might wonder if perhaps the $a_0$ scaling depends on the way the field goes to zero. For this reason we have also considered $a(\phi)=a_0(1+\cos\phi)^2$ for $|\phi|<\pi$ and $a(\phi)=0$ for $|\phi|>\pi$, which has a different decay at $\phi=\pm\pi$. In one part of the calculation of NLO one finds that the dominant contribution comes from a region of the $\phi$ variables that scales differently in $a_0$ compared to the first example. However, we still find the same form as in~\eqref{compactExample1}; the only difference is the numerical coefficients.

\subsection{Perturbation theory}\label{perturbationSection}

\begin{figure}
\includegraphics[width=\linewidth]{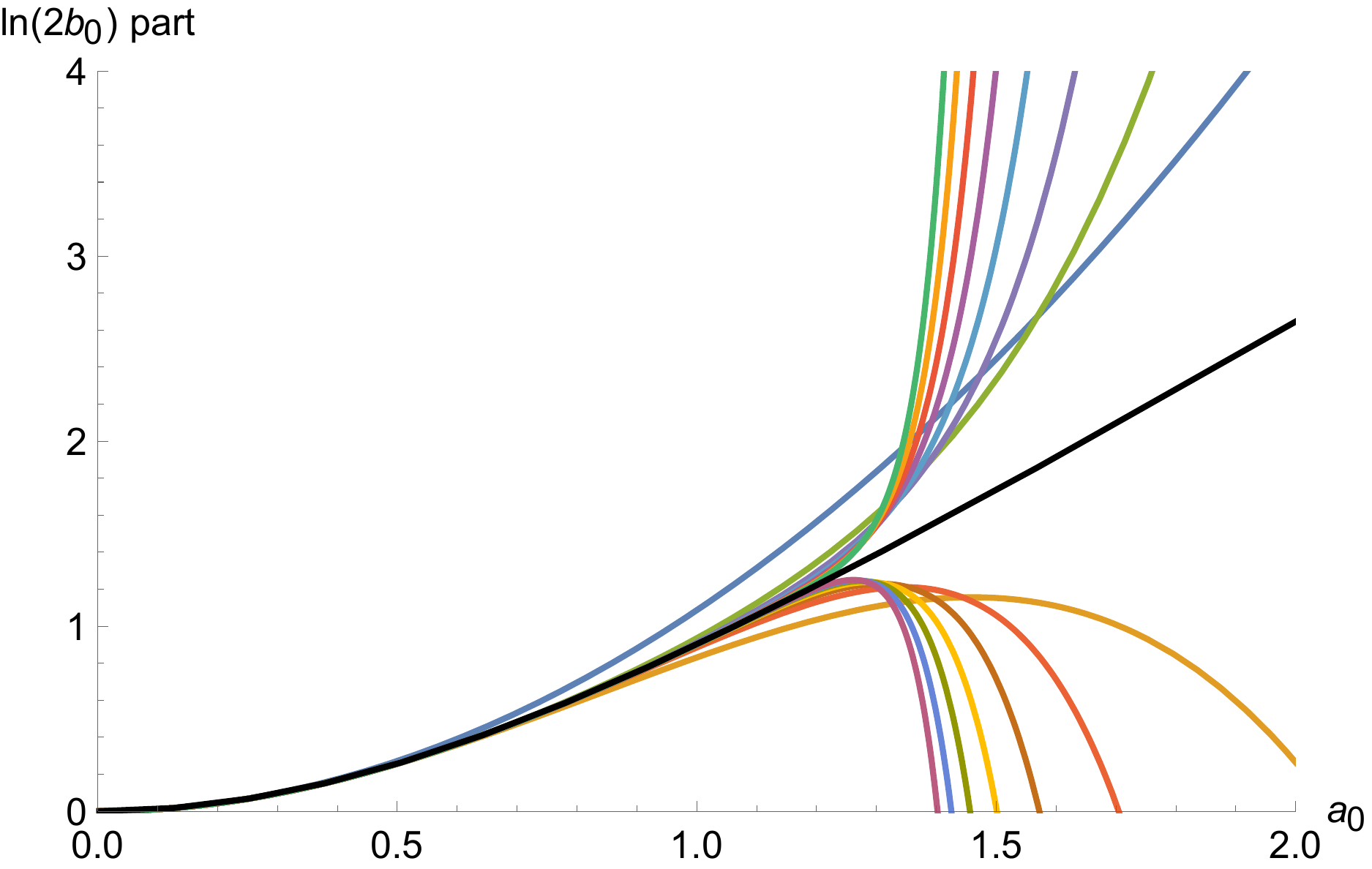}
\includegraphics[width=\linewidth]{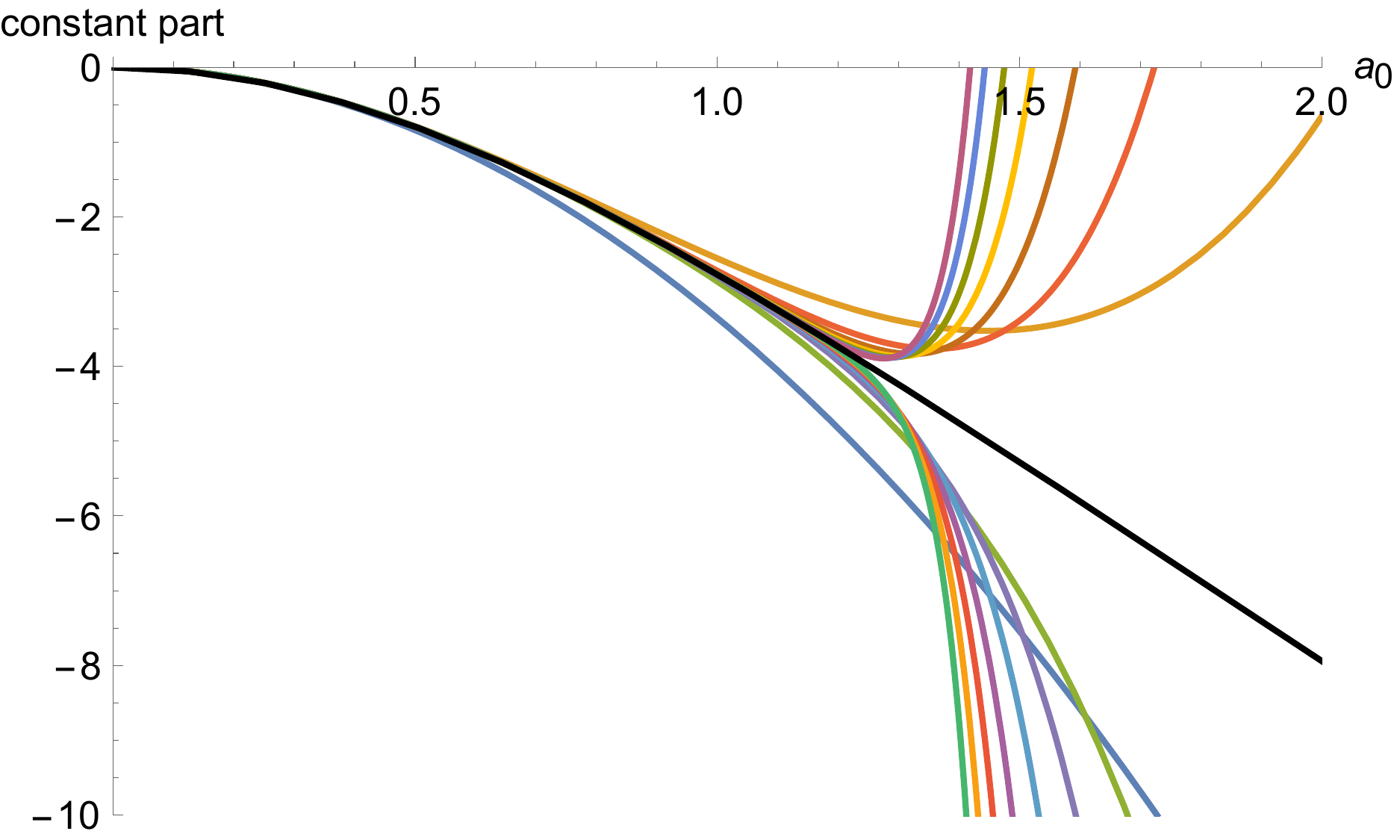}
\caption{Exact evaluation of~\eqref{totalPgeneral} (black lines) compared to the perturbative sums including 1 and up to 15 of the first terms in the $a_0^2$ expansion.}
\label{perturbativeSumsFig}
\end{figure}

In the previous two sections we studied the large-$a_0$ expansion and showed that by including NLO we obtain approximations that are good all the way down to $a_0\gtrsim2$. In this section we study the small-$a_0$ expansion. In contrast to the large-$a_0$ expansion, it is quite straightforward to obtain a perturbation series in $a_0$. We do not need to figure out how to rescale integration variables, we just have to expand the original integrand in a power series in $a_0$ and then perform the integrals at each order numerically. We consider again the compact, short field in~\eqref{compactField}. We find
\be\label{PLC}
\mathbb{P}=\alpha^2[L(a_0)\ln(2b_0)+C(a_0)] \;,
\ee       
where $L(a_0)$ and $C(a_0)$ can be expanded in a power series in $a_0^2$ with coefficients with alternating sign and decreasing in absolute value,
\be\label{Lexpansion}
L=1.1a_0^2-0.26a_0^4+...-0.00022a_0^{28}+0.00014a_0^{30}-...
\ee
\be\label{Cexpansion}
C=-3.6a_0^2+0.80a_0^4-...+0.00050a_0^{28}-0.00031a_0^{30}+...
\ee
The ratios of neighboring coefficients seem to converge to $c_{n}/c_{n-1}\sim-0.64$, indicating a finite radius of convergence of $a_0^2\sim 1.5$. We can check this from the zeros of the effective mass at imaginary $a_0$: In the denominator of the integrand in~\eqref{totalPgeneral} we have $M^2=1+a_0^2[(M^2-1)/a_0^2]$. The maximum of $[(M^2-1)/a_0^2]$ is $\approx0.76$ (reached in the $\mathcal{P}_{12}$ region), which means that the singularity closest to the origin is at $a_0^2\sim-1.5$, so the radius of convergence is $a_0^2\sim 1.5$.
This is also agrees with Fig.~\ref{perturbativeSumsFig}, where we compare an exact evaluation of~\eqref{totalPgeneral} with the perturbation series. Fig.~\ref{perturbativeSumsFig} shows that at $a_0\sim1$ one can still obtain better precision by including more terms, but at $a_0\sim1.2$ the perturbative sums deviate from the exact result regardless of how many terms one adds. (The radius of convergence for other processes in different regimes has been studied in~\cite{Reiss:1980zz,ReissConvergentPerturbation,ReissSLACnonPerturbative}.)  

\begin{figure}
\includegraphics[width=\linewidth]{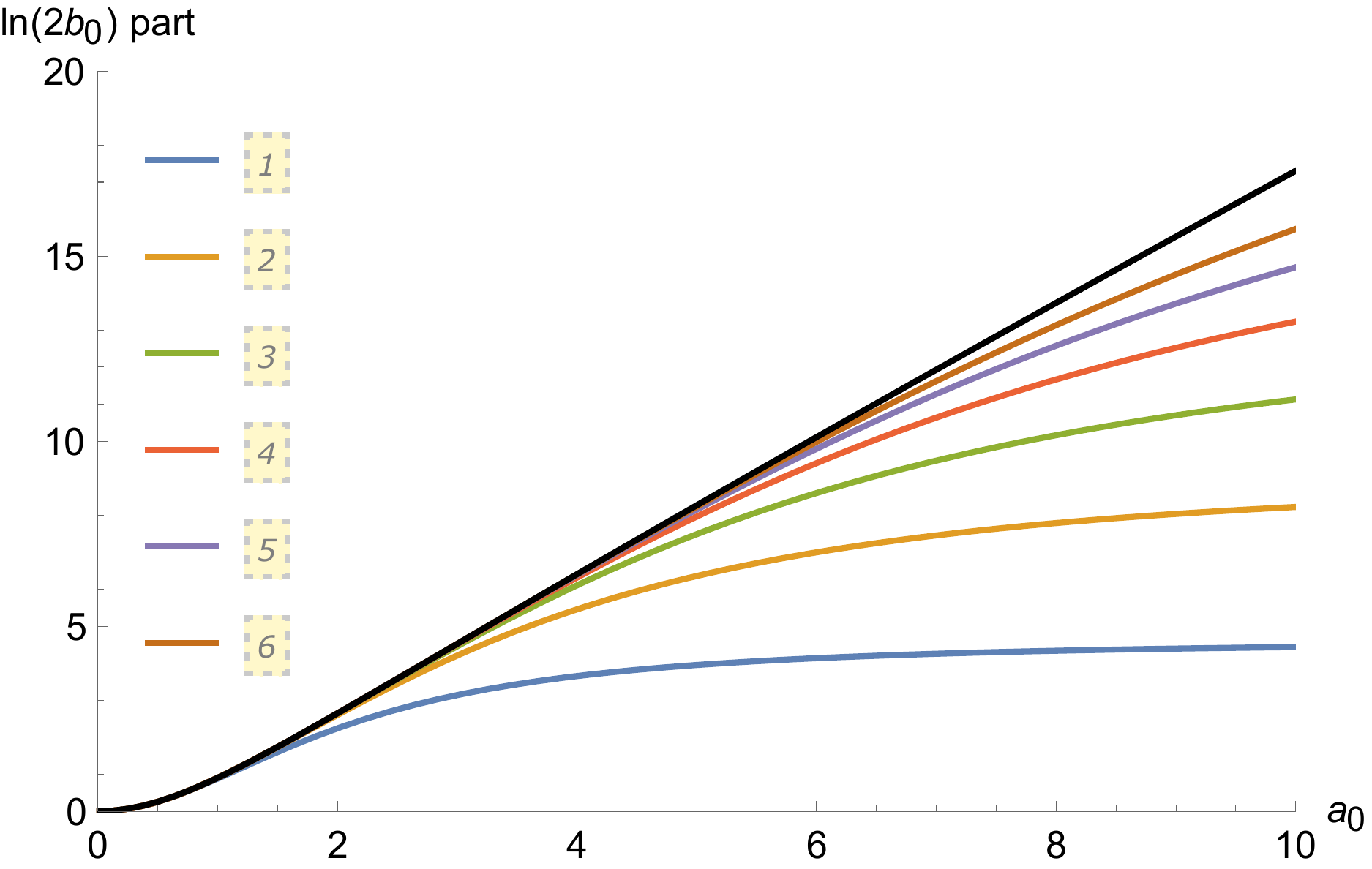}
\includegraphics[width=\linewidth]{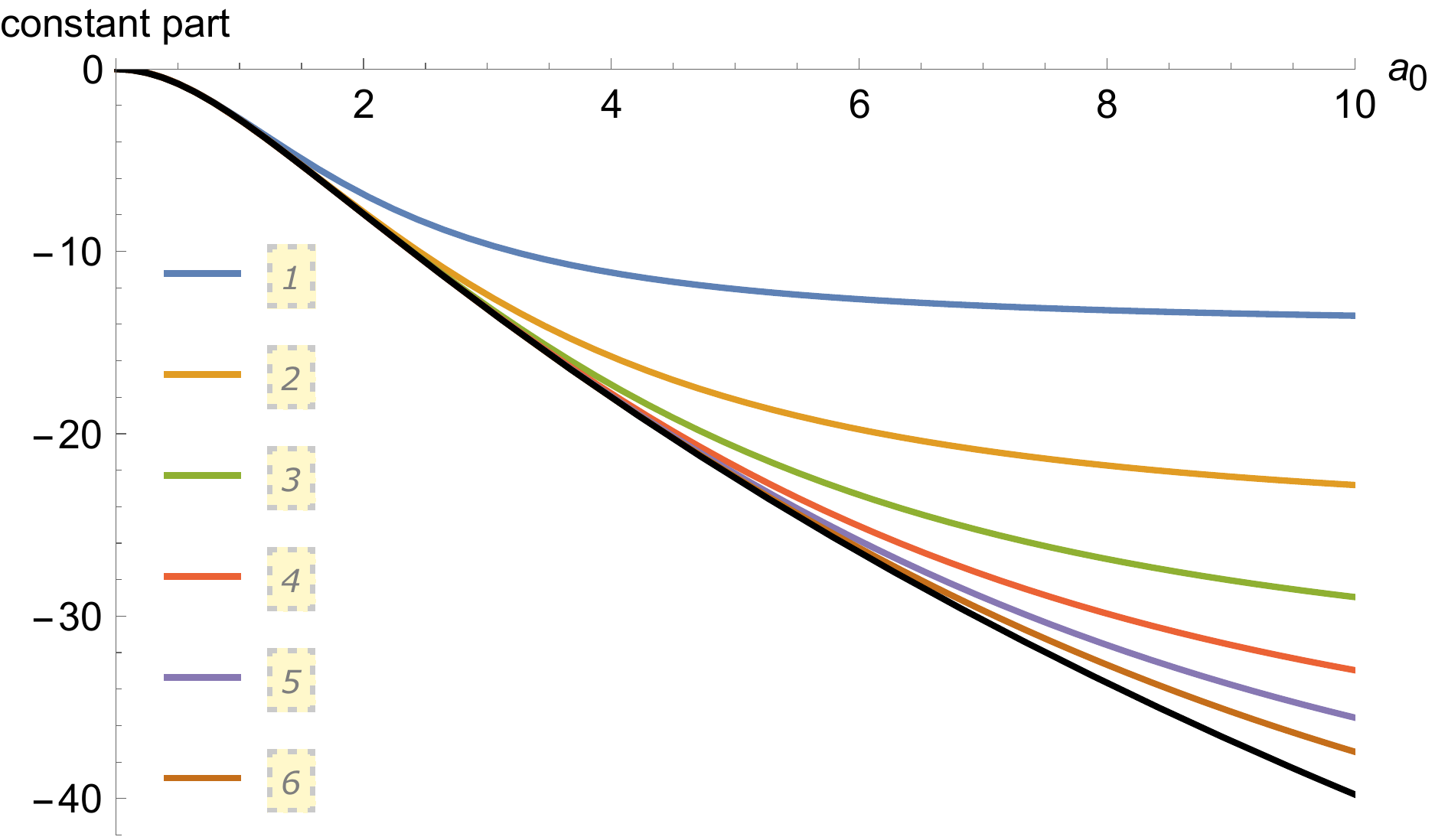}
\caption{The black lines are exact evaluation of~\eqref{totalPgeneral} and the other lines show the Pad\'e approximants with $N=M=1,2,...,6$.}
\label{PadeFig}
\end{figure}

\begin{figure}
\includegraphics[width=\linewidth]{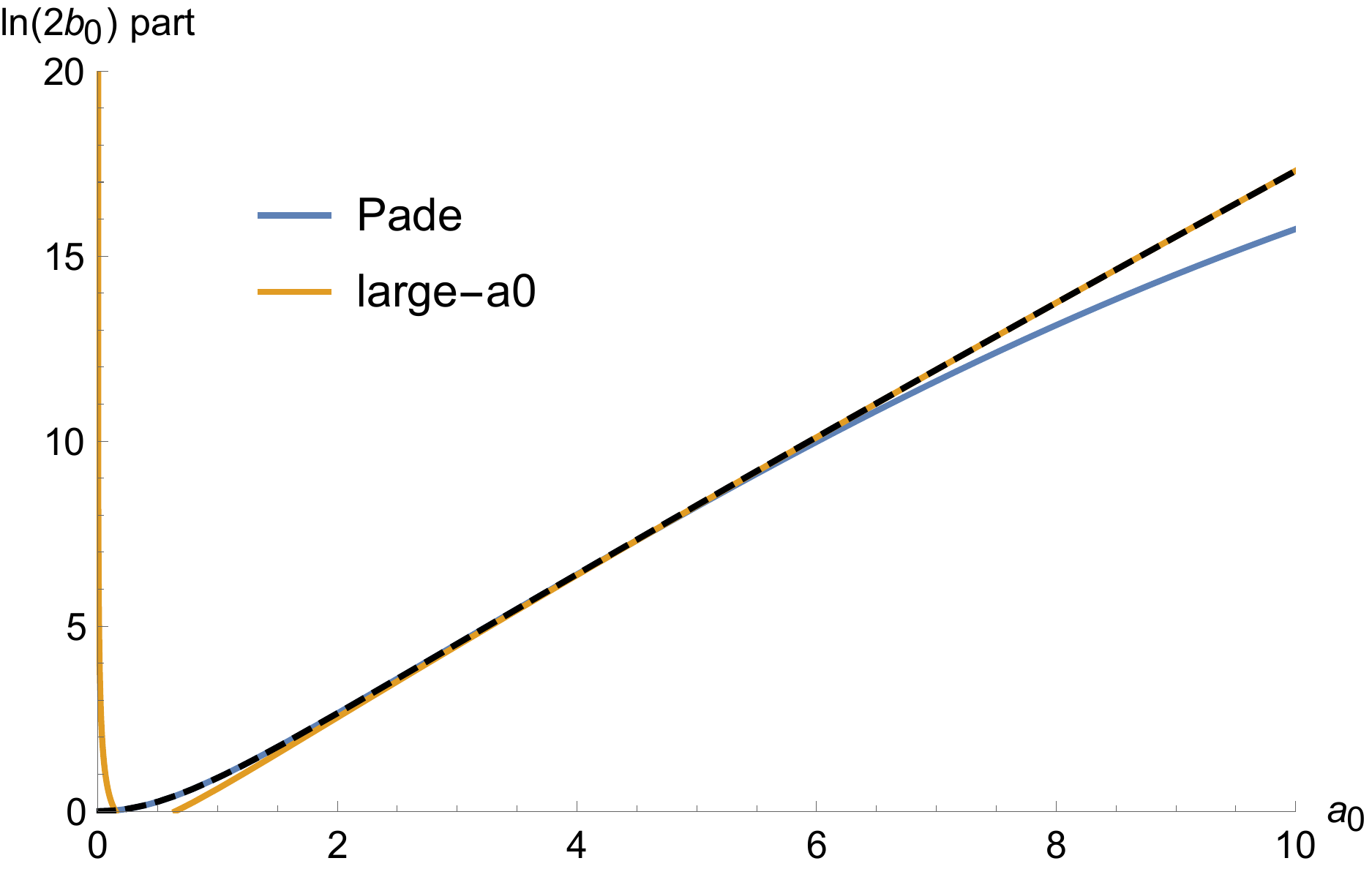}
\includegraphics[width=\linewidth]{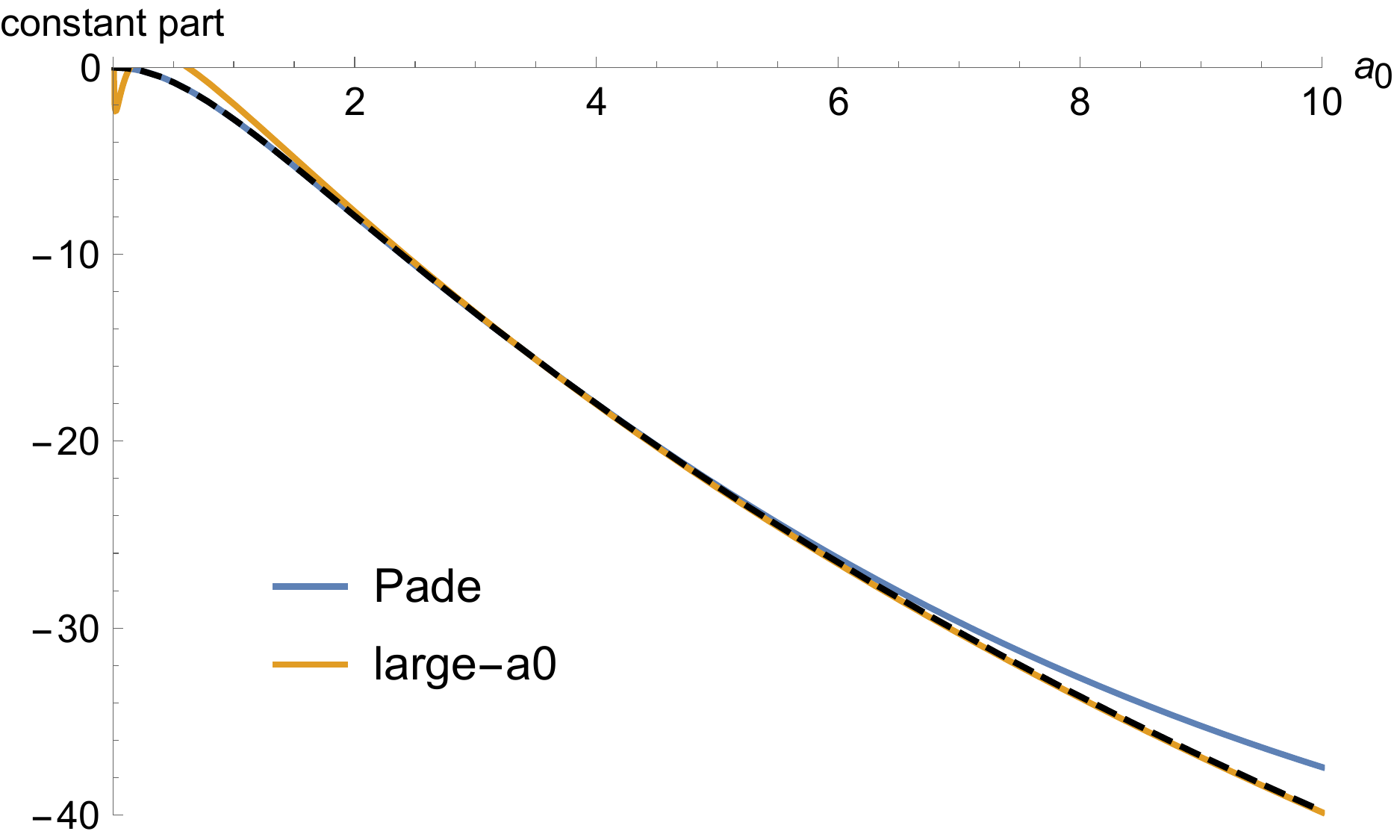}
\caption{The black dashed lines are exact evaluation of~\eqref{totalPgeneral} and the other lines show the Pad\'e approximant with $N=M=6$ and the large-$a_0$ expansion in~\eqref{compactExample1}.}
\label{PadeVsLargeFig}
\end{figure}

So, perturbation theory seems to be limited to small $a_0$ ($\lesssim1.2$ in this example), which is what one might have expected. However, there is growing interest in the field of extracting information encoded in perturbation series (around the origin in this case) to study different regions of parameter space, see e.g.~\cite{Costin:2019xql,Costin:2020hwg,Florio:2019hzn,Caliceti:2007ra} and references therein. In our case we have resummed the perturbation series into Pad\'e approximants~\cite{Baker1961,BenderOrszag,KleinertPhi4,ZinnJustinBook},
\be
P_M^N(a_0^2)=\frac{\sum_{n=0}^N A_na_0^{2n}}{\sum_{n=0}^M B_na_0^{2n}} \;,
\ee
where $B_0=1$, (in our case) $A_0=0$, and the other coefficients are obtained by expanding into a perturbation series and matching with~\eqref{Lexpansion} and~\eqref{Cexpansion}. 
Pad\'e approximants are sometimes used together with Borel resummation in order to treat asymptotic perturbation series~\cite{Costin:2019xql,Costin:2020hwg}, see \cite{Florio:2019hzn} for an application to the Euler-Heisenberg effective action and Schwinger pair production. However, in this case we have a convergent series, so we apply the Pad\'e method directly.  
In Fig.~\ref{PadeFig} we compare the first few (diagonal) approximants with $N=M$. These Pad\'e approximants give a good agreement with the exact result beyond the radius of convergence all the way up to $a_0\sim7$. So, even though we cannot use a direct sum of the perturbation series at $1.5\lesssim a_0\lesssim 10$, the behavior of the probability in this region is encoded in the coefficients of the perturbation series around $a_0=0$. Since we showed in the previous sections that the large-$a_0$ expansion (with NLO included) is good all the way down to $a_0\gtrsim2$, we now have a significant overlap where the small- and large-$a_0$ expansions give numerically basically the same results, as illustrated in Fig.~\ref{PadeVsLargeFig}. This means that we have analytical approximations for any value of $a_0$.

\begin{figure}
\includegraphics[width=\linewidth]{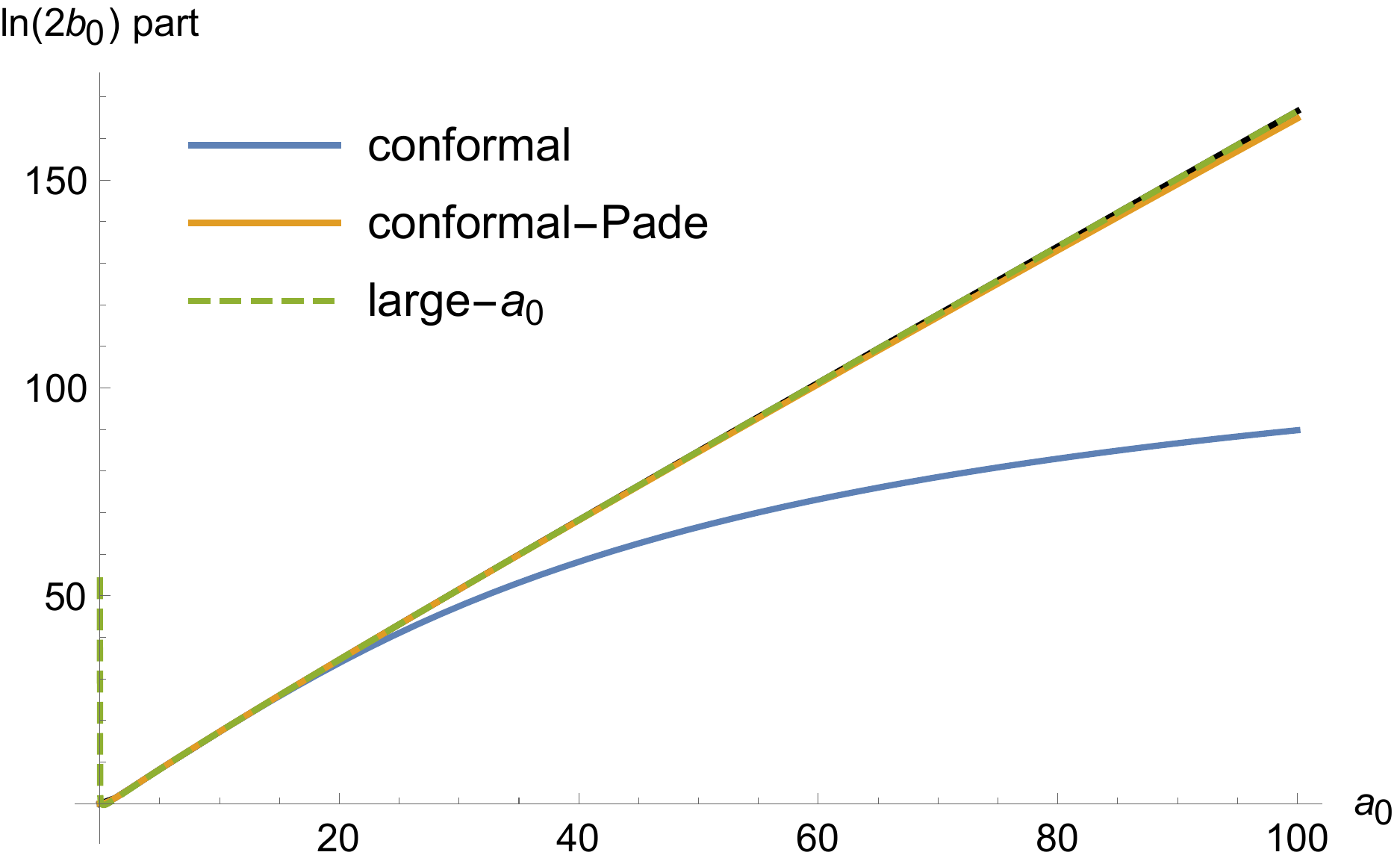}
\includegraphics[width=\linewidth]{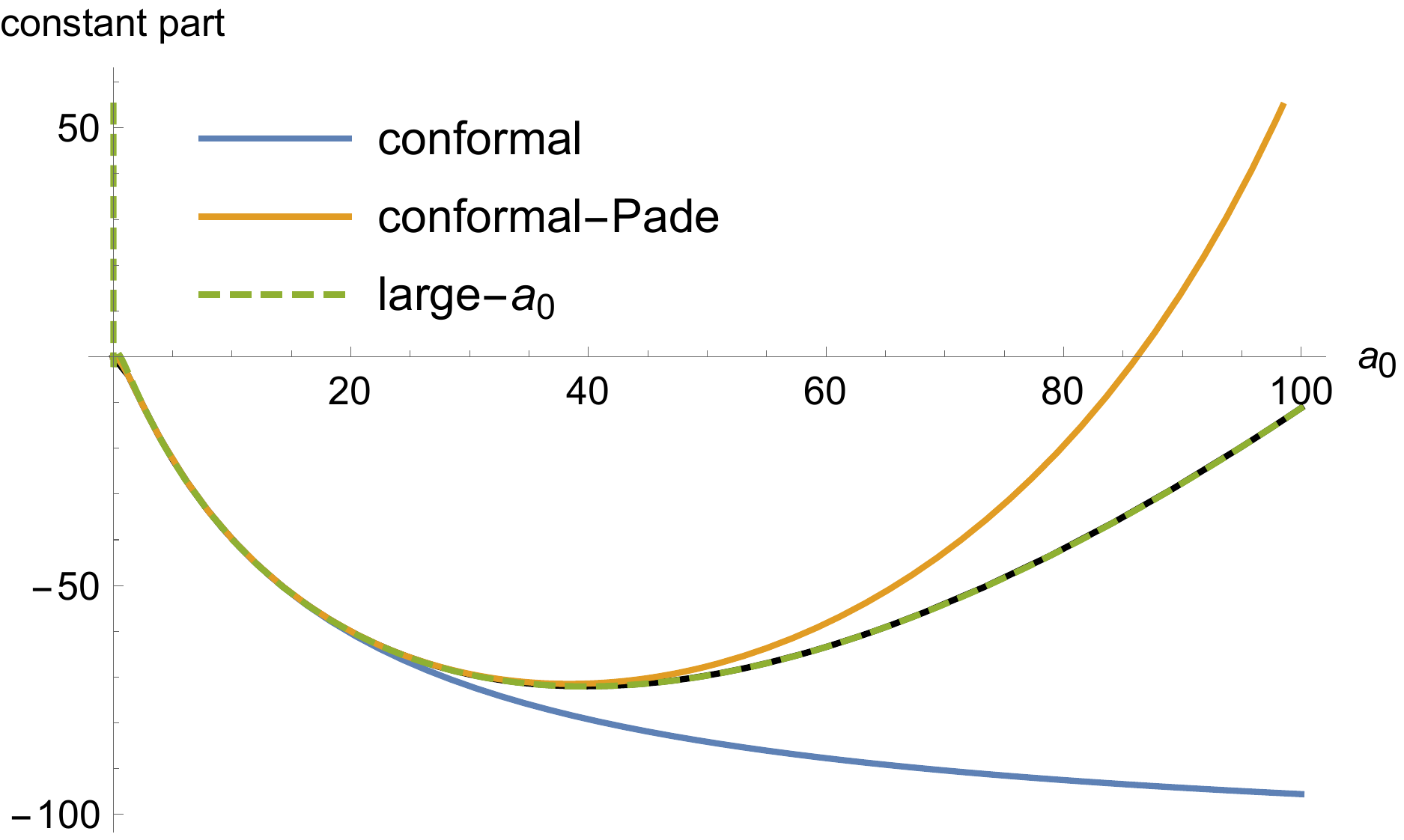}
\caption{The black lines are exact evaluation of~\eqref{totalPgeneral} and the other lines show: the perturbative sum in the conformal variable $z$ up to $z^{30}$; the Pad\'e approximant of the conformal series with $N=M=10$ ($N=M=14$) for the log part (constant part); and the large-$a_0$ expansion in~\eqref{compactExample1}.}
\label{ConformalFig}
\end{figure}

We can extend the reach of perturbation theory further using a conformal map~\cite{KleinertPhi4,ZinnJustinBook,Costin:2019xql,Costin:2020hwg,Florio:2019hzn,Guillou1980},
\be
z=\frac{\sqrt{1+\frac{a_0^2}{a_s^2}}-1}{\sqrt{1+\frac{a_0^2}{a_s^2}}+1} \qquad a_0^2=\frac{4a_s^2z}{(1-z)^2} \;,
\ee
which maps the singularity at $a_0^2=-a_s^2$ to the unit circle\footnote{The first singularity is at $a_0^2\sim-1.5$. In the plots we have chosen $a_s$ to be the (numerically obtained) value of the first singularity, as is usually done. This is expected to be the optimal conformal map~\cite{optimalConformal}. However, in this case this only gives a slight improvement compared to simply setting e.g. $a_s\to1$. In this particular case, this can be understood by noting that, while the singularity closest to the origin is determined by $\mathcal{P}_{12}$, the dominant contribution at large $a_0$ comes from $\mathcal{P}_{22}$.}. Instead of expanding in a power series in $a_0^2$, we expand in $z$. Using only the conformal map also allows us to go beyond the radius of convergence; for the example in Fig.~\ref{ConformalFig} we find agreement up to $a_0\sim 20$ by including terms up to $z^{30}$. We can reach further if we perform a Pad\'e resummation of the conformal series; in Fig.~\ref{ConformalFig} we find agreement up to $a_0\sim50$ with a Pad\'e approximant with $N=M=14$ (for the constant part; the log part is much better).  

So, by calculating a sufficient number of the coefficients in the perturbation series around $a_0=0$, one can obtain a good approximation even for large $a_0$. We are therefore led to consider large orders. A direct numerical integration can become challenging if we need to go to very high orders, but at sufficiently high orders we can use a semi-analytical approach. At $\mathcal{O}(a_0^{2N})$, the most important part of the integrand is given by
\be
\left(-\frac{M^2-1}{a_0^2}\right)^N \;,
\ee
which makes the integrand sharply peaked at the point where $M^2$ is at maximum. By exponentiating this factor as (cf.~\cite{Lam:1968tk})
\be
\exp\left\{N\ln\left[\frac{M^2-1}{a_0^2}\right]\right\} \;,
\ee
we can use saddle-point methods to perform the integrals\footnote{Strictly speaking, the maximum is not necessarily a saddle point. For example, in the contribution from $\mathcal{P}_{22}$ the maximum occurs on the boundary of the integration region, which means that, after a suitable choice of integration variables, we have one integration variable with linear rather than quadratic fluctuation around the maximum.}
To leading order the rest of the integrand is simply evaluated at the maximum. Let $L_{ijN}$ and $C_{ijN}$ be the contributions from region $\mathcal{P}_{ij}$ to the coefficients of $L$ and $C$ in~\eqref{PLC} at $\mathcal{O}(a_0^{2N})$. To leading order we find
\be
L_{12N}\approx-\frac{0.22(-0.67)^N}{N}
\ee   
\be
L_{22N}\approx-\frac{0.35(-0.54)^N}{N^\frac{3}{2}}
\ee
\be
L_{13N}\approx-\frac{0.015(-0.56)^N}{N^\frac{1}{2}}
\ee
\be
C_{12N}\approx-\frac{(0.11\ln N-0.81)(-0.67)^N}{N}
\ee 
\be
C_{22N}\approx-\frac{(0.17\ln N-1.1)(-0.54)^N}{N^\frac{3}{2}}
\ee
\be
C_{13N}\approx-\frac{(0.0075\ln N-0.056)(-0.56)^N}{N^\frac{1}{2}} \;.
\ee
From these large-order approximations we see explicitly that we have convergent series. The ratio test gives of course the same radius of convergence as we found above.

\section{Heavy mass and low energy}\label{saddlePointSection}

\begin{figure}
\includegraphics[width=\linewidth]{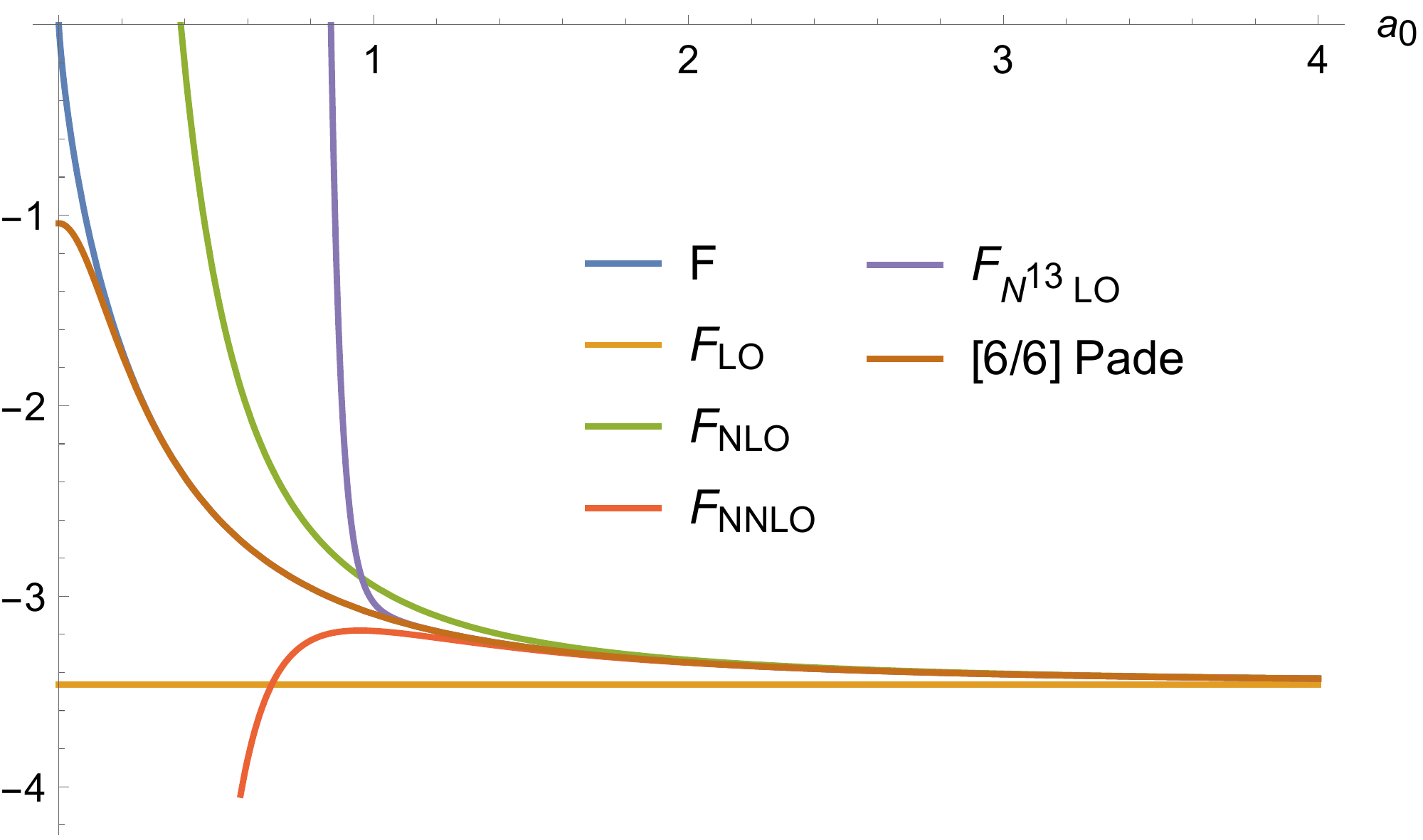}
\includegraphics[width=\linewidth]{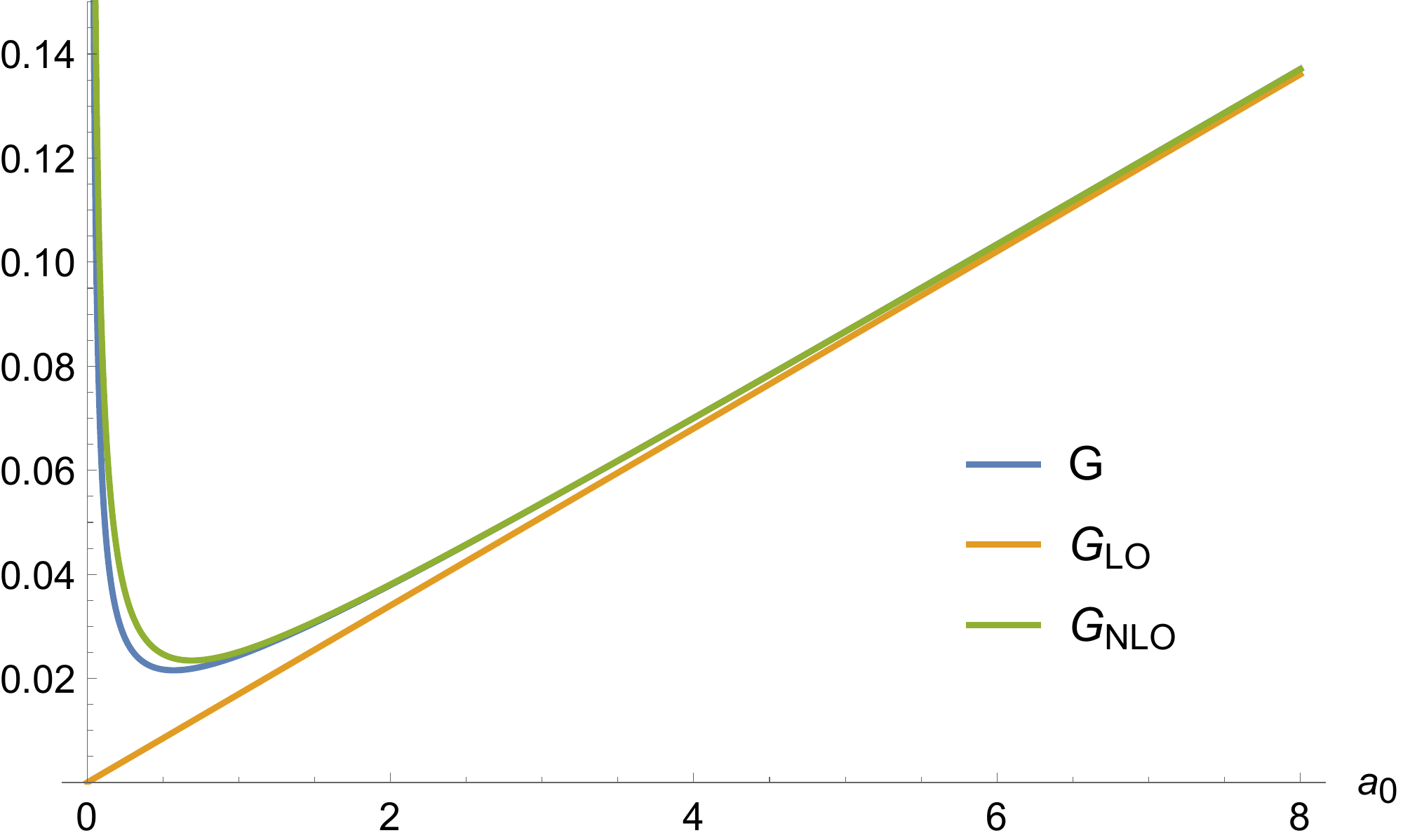}
\caption{The exponent and the prefactor for $\mu\to\infty$, low-energy and a sinusoidal field. The probability is given by $\mathbb{P}=\alpha^2E^2Ge^{F/E}$, where $E$ is the field strength. ``LO'', ``NLO'', ``NNLO'' and ``$\text{N}^{13}$LO'' are obtained by expanding in $1/a_0$. ``$\text{N}^{13}$LO'' includes terms up to $1/a_0^{26}$, and $[6/6]$~Pad\'e is the $N=M=6$ Pad\'e approximant for the perturbative series in $1/a_0^2$.}
\label{sinCompSaddleFig}
\end{figure}

In the previous section we saw that, in the high-energy limit, the ``lightfront-time-instantaneous'' $\mathbb{P}_{11}$ contributes to leading order and is only smaller than the dominant term by a logarithmic term $\ln b_0$. This is interesting because $\mathbb{P}_{11}$ gives in general only a small contribution. In this section we study a regime where $\mathbb{P}_{11}$ alone gives the dominant contribution. 

So, consider the limit where the mass of the initial particle is much heavier than the mass of the pair, $\mu\to\infty$. This is a relevant limit since already a muon is much heavier than an electron. 
Our calculations could also be relevant for processes involving millicharged particles, see e.g.~\cite{Gies:2006hv,MillichargedPresentations}, but here we will focus on electrons and muons.
In this limit the momentum of the initial particle does not change much during the process. So, we change variables from $s_2=(1-s_1)u$ to $u$ and from $s_1=1/(1+t)$ to $t$, rescale $t\to x/\mu$ and then take the limit $\mu\to\infty$. For $\mathbb{P}_{22}$ the lightfront-time integrals for the first step become free, i.e. the background field enters only via the second step. We can therefore perform the $\sigma_{21}$ and $\theta_{21}$ integrals. The first term in~\eqref{StepsForStepsSep} vanishes because the $\theta_{21}$ contour can be closed in the upper half of the complex plane and there are no poles there. This means that the two-step part $\mathbb{P}_{\rm two}$ does not contribute in the limit where the initial particle is much heavier than the pair.
For the second term the $\sigma_{21}$ integral is trivial and gives
\be
-\frac{|\theta_{43}-\theta_{21}|}{2}=\int\frac{\ud r}{2\pi}\frac{e^{i(\theta_{43}-\theta_{21})r}-1}{r^2} \;.
\ee 
The $\theta_{21}$ integral can now be performed with the residue theorem, and then the $r$ integral can be expressed in terms of an incomplete Gamma function.
We recover exactly~\eqref{PExactCoulomb}, so the infinite mass limit of muon trident agrees with pair production by the superposition of a Coulomb field and a plane wave, as expected. 

The high-energy limit therefore just reproduces the result in the previous section, so we consider now instead the low-energy limit. We have $b_0/\mu=\omega$ and we consider $\omega\ll1$. In this regime we can perform all the integrals with the saddle-point method. For the momentum integrals we have a saddle point at $u=1/2$ and $x=2M(\sigma_{43},\theta_{43})$. We find
\be
\mathbb{P}_{11}=-\frac{i\alpha^2\omega}{4\pi}\int\frac{\ud\sigma\ud\theta}{\theta^3M}\exp\left\{\frac{2i\theta M}{\omega}\right\}
\ee 
and
\be
\mathbb{P}_{22}=\frac{\alpha^2\omega^2}{8\pi}\int\ud\sigma\ud\theta\frac{2+D_2}{\theta^4M^4}\exp\left\{\frac{2i\theta M}{\omega}\right\} \;.
\ee
Although one could perform the remaining integrals numerically, that would not give a more accurate result than performing them with the saddle-point method, because we have already performed the momentum integrals with the saddle-point method. However, we can already see that $\mathbb{P}_{11}$ gives the dominant contribution for arbitrary field shape, as $\mathbb{P}_{22}\sim\omega \mathbb{P}_{11}\ll \mathbb{P}_{11}$. Contrast this with the case where all the masses are equal, where the opposite is true~\cite{Dinu:2017uoj}, i.e. $\mathbb{P}_{11}\sim\chi\mathbb{P}_{22\to1}\ll \mathbb{P}_{22\to1}$.
Thus, here the lightfront instantaneous term $\mathbb{P}_{11}$ gives the dominant contribution.

Note also that the saddle points for the lightfront-time integral are determined by the function $\theta M$, while in the equal-mass case as well as other processes such as nonlinear Compton scattering, Breit-Wheeler pair production or double nonlinear Compton scattering, the saddle points are instead determined by $\theta M^2$~\cite{Dinu:2018efz}. In~\cite{Dinu:2018efz} we found explicit saddle-point approximations for an entire class of field shapes for processes with $\theta M^2$ in the exponent. However, these fields lead to transcendental equations for the saddle points here. It is, of course, still simple to obtain the saddle points numerically, expand around the saddle point and perform the resulting Gaussian integral analytically. 
For $a_0\gg1$ we can still find fully analytical results. We have $M^2=1+a_0^2\dot{f}(\sigma)\theta^2/12$ and a saddle point at $\theta=i\sqrt{6}/(|a_0\dot{f}(\sigma)|)$, and we find
\be\label{P11domLCF}
\mathbb{P}\approx\mathbb{P}_{11}\approx\frac{\alpha^2}{2\sqrt{\pi}}\int\ud x^\LCp\left(\frac{|E(x^\LCp)|}{2\sqrt{3}}\right)^\frac{5}{2}\exp\left\{-\frac{2\sqrt{3}}{|E(x^\LCp)|}\right\} \;,
\ee    
where $E(x^\LCp)=\omega a'(\sigma/\omega)$. For a constant field we recover Eq.~(44) in~\cite{Ritus:1972nf}.

We can also perform the remaining integral with the saddle-point method. 
If $a_0$ is only moderately large we need to include higher-order corrections to the leading order~\eqref{P11domLCF}. By expanding in $1/a_0$ we obtain
\be\label{PFG}
\mathbb{P}\approx\alpha^2E^2Ge^{F/E} \;,
\ee
where $E=E(0)$ is the field maximum,
\be
\begin{split}
F\approx-2\sqrt{3}&\left(1-\frac{3\zeta}{20a_0^2} \right. \\
&\left.+\frac{3\zeta^2}{5600a_0^4}\left[143-15\frac{f^{(5)}_0}{\zeta^2}-21\frac{[f^{(4)}_0]^2}{\zeta^3}\right]\right) \;,
\end{split}
\ee
and
\be
G\approx\frac{a_0}{24\sqrt{6\zeta}}\left(1+\frac{\zeta}{40a_0^2}\left[22-3\frac{f^{(5)}_0}{\zeta^2}-3\frac{[f^{(4)}_0]^2}{\zeta^3}\right]\right) \;,
\ee
where $\zeta=-f^{(3)}_0>0$ and $f^{(n)}_0=\partial_\phi^n|_0f$. We have included one more order in the exponential part because even a small difference from the exact $F$ can lead to a non-negligible difference in $\mathbb{P}$ due to the factor of $1/E\gg1$ in the exponent.
Note that all terms are local, they come from the region where the two lightfront time variables are close, which is seen from the fact that they are expressed in terms of derivatives of the field (evaluated at the maximum). This is what one can expect in a LCF expansion in $1/a_0$, but contrast this with the high-energy limit in the previous section, where the next-to-leading order corrections are nonlocal. Note also that we do not automatically have a local expression just because we can perform the lightfront time integrals with the saddle point method (which we can do as long as $E$ is small enough), because, although the average variable $\sigma=(\phi_2+\phi_1)/2$ is in general forced to be close to the field maximum, for $a_0\sim1$ we have a saddle point at $\theta=\phi_2-\phi_1\sim i$, so in that case the imaginary part of $\phi_2$ and $\phi_1$ do not have to be close, i.e. the result would be nonlocal.


In order to compare this expansion with the exact result, we consider a linearly polarized monochromatic field, $a(\phi)=a_0\sin\phi$. We have a saddle point with $\sigma=0$ and $\theta$ determined by\footnote{For comparison we note that for a circularly polarized monochromatic field the saddle point is determined by $\text{sinc}\,\theta=1+\frac{1}{a_0^2}$.}
\be
1+\frac{2}{a_0^2}-\frac{\cos\theta}{2}-\frac{\sin\theta}{2\theta}=0 \;.
\ee
For $a_0\gg1$ we have $\theta=i\sqrt{6}/a_0$, for $a_0\ll1$ $\theta=i\ln(8/a_0^2)$. 
For $a_0\ll1$ the exponential part becomes
\be
\mathbb{P}\sim a_0^\frac{4}{\omega} \;,
\ee
which is the expected perturbative result since $2/\omega$ photons have to be absorbed to produce the pair in the limit where the initial particle is very heavy.
For $a_0\sim1$ we can solve the saddle-point equation numerically. (The corresponding equation in the equal-mass case can be solved explicitly in terms of an inverse trigonometric function.) 
The result is compared in Fig.~\ref{sinCompSaddleFig} with the corresponding approximation~\eqref{PFG}. We see that by including the first couple of terms in the $1/a_0$ expansion we obtain a good approximation already for $a_0\gtrsim1.5$.
It is straightforward to obtain higher orders in $1/a_0^2$, but, as we see in Fig.~\ref{sinCompSaddleFig} (where we include terms up to $1/a_0^{26}$), there is a limit for how low $a_0$ that can be reached with a direct sum of the perturbation series in $1/a_0^2$. However, by resumming this series into a Pad\'e approximant, we can reach much lower $a_0$. So, we see that resummation methods can be useful for perturbation series in both $a_0$ and $1/a_0$.

\subsection{Production of a muon pair}

We can also consider the process where the initial particle is much lighter than the pair, for example an electron producing a muon pair. If $a_0=E/(m_e\omega)\gg1$ it could still be that the muon nonlinearity parameter $a_\mu=E/(m_\mu\omega)$ is not large. Since the muon is much heavier than the electron $\mu\approx207$ one can expect an exponential suppression, so we perform the integrals with the saddle-point method. We consider for simplicity a Sauter pulse $f(\phi)=\tanh\phi$. For the two-step we find
\be
\mathbb{P}_{\rm two}\sim\exp\left\{-\frac{4}{\chi}\left(\mu^3a_\mu\Lambda+\sqrt{\frac{2\mu^3a_\mu\Lambda}{3}}\right)\right\} \;,
\ee
where $\Lambda=(1+a_\mu^2)\text{arccot}a_\mu-a_\mu$. For $a_\mu\gg1$ this reduces to
\be
\mathbb{P}_{\rm two}\sim\exp\left\{-\frac{8}{3\chi}(\mu^3+\mu^{3/2})\right\} \;,
\ee
which agrees with Eq.~(24) in~\cite{Ritus:1972nf}. However, the one-step scales as
\be
\mathbb{P}_{11}\sim\exp\left\{-\frac{4\sqrt{3}\mu^2}{\chi}+\frac{2\sqrt{3}}{\chi}\left(1+\frac{6}{5a_\mu^2}\right)\right\}
\ee 
and is therefore exponentially larger than the two-step. 
For $a_\mu\gg1$ we can neglect the third term. The first, leading term, $-\frac{4\sqrt{3}\mu^2}{\chi}$ agrees with the result in~\cite{Baier}. 
So, this is another regime where the one-step dominates over the two-step. 
However, for the production of a muon pair by an electron we have $4\sqrt{3}\mu^2\approx3\times10^{5}$, so $\chi$ would have to be very large for this to not be completely negligible. 
It could therefore be more interesting to consider the opposite process, where there is a muon in the initial state with $a_\mu\sim1$ (which means $a_0\gg1$). We leave this for future studies.

\section{Large $\chi$ from small-$\chi$ expansion}\label{resumchi}

In this section we will study the $\chi$ dependence of the LCF result. In particular, we will show how asymptotic (divergent) power series in $\chi$ can be resummed using Borel-Pad\'e-conformal methods~\cite{Guillou1980,Caliceti:2007ra,Costin:2019xql,Costin:2020hwg,KleinertPhi4,ZinnJustinBook} to obtain a good approximation up to very large $\chi$.

\subsection{Nonlinear Breit-Wheeler pair production}

We start for simplicity with nonlinear Breit-Wheeler pair production. In LCF the probability is given by (see e.g.~\cite{Dinu:2017uoj})
\be
\mathbb{P}=\alpha a_0\int\ud\sigma R \;,
\ee  
where
\be\label{RfromAiry}
R=\int\ud s\left(\text{Ai}_1(\xi)-\kappa\frac{\text{Ai}'(\xi)}{\xi}\right) \qquad \xi=\left(\frac{r}{\chi_\gamma}\right)^\frac{2}{3}
\ee
and
\be
r=\frac{1}{s(1-s)} \qquad
\kappa=\frac{s}{1-s}+\frac{1-s}{s} \;.
\ee
We use $\chi_\gamma=a_0kl$, where $l_\mu$ is the momentum of the incoming photon, to distinguish it from the electron $\chi$ in trident. $s=kp'/kl$ is the fraction of the longitudinal momentum given to the produced electron. $\text{Ai}(\xi)$ is the Airy function and
\be
\text{Ai}_1(\xi)=\int_\xi^\infty\ud t\text{Ai}(t) \;.
\ee
We could consider a field with a locally constant $\chi_\gamma(\sigma)$, but here we focus on the integrand $R$ at a given value of $\chi_\gamma$. 
For small and large $\chi_\gamma$ the probability is given by~\cite{Nikishov1}
\be\label{LCFBWlow}
\chi_\gamma\ll1: \qquad R=\frac{3}{16}\sqrt{\frac{3}{2}}\exp\left\{-\frac{8}{3\chi_\gamma}\right\}
\ee
and
\be\label{LCFBWhigh}
\chi_\gamma\gg1: \qquad R=\frac{15\times3^\frac{2}{3}\Gamma^4\left[\frac{2}{3}\right]}{28\pi^2\chi_\gamma^\frac{1}{3}} \;.
\ee
The goal now is to obtain sufficiently many higher-order corrections to~\eqref{LCFBWlow} in order to make a resummation that works up to $\chi$ large enough so that we have agreement with~\eqref{LCFBWhigh}. We can do this by first expanding the Airy functions at large arguments, 
\be
\text{Ai}_1(\xi)=\frac{1}{2\sqrt{\pi x}}\left(1-\frac{41}{48x}+\frac{9241}{4608x^2}+\dots\right)\exp\left\{-\frac{2x}{3}\right\} \;,
\ee 
where $x=\xi^\frac{3}{2}=r/\chi_\gamma$, and similarly for $\text{Ai}'(\xi)/\xi$. The $s$ integral can now be performed by expanding the integrand around the saddle point at $s=1/2$\footnote{Resummations of saddle-point expansions have been discussed in~\cite{Serone:2017nmd}.}. We find
\be
R=\frac{3}{16}\sqrt{\frac{3}{2}}T\exp\left\{-\frac{8}{3\chi_\gamma}\right\} \;,
\ee
where 
\be\label{Tdefinition}
T=\sum_{n=0}^\infty T_n\chi_\gamma^n=1-\frac{11}{64}\chi_\gamma+\frac{7985}{73728}\chi_\gamma^2-\frac{4806425}{42467328}\chi_\gamma^3+\dots
\ee
We have calculated the first 56 terms, but it is not difficult or time consuming to obtain more terms. By plotting the ratio of neighboring coefficients $T_n/T_{n-1}$, it is clear that they grow factorially. It is therefore natural to make a Borel transform
\be\label{BorelBWdefinition}
{\rm BT}(t)=\sum_{n=0}^\infty\frac{T_n t^n}{n!} \;.
\ee
We have a finite number of terms for ${\rm BT}$. We resum this truncated series into a Pad\'e approximant, ${\rm PBT}(t)$, which gives a ratio of two polynomial functions of $t$. The final result is now obtained by a Laplace transform
\be\label{LaplaceTransform}
T_{\rm re}(\chi_\gamma)=\int_0^\infty\frac{\ud t}{\chi_\gamma}e^{-t/\chi_\gamma}{\rm PBT}(t) \;.
\ee 

\begin{figure}
\includegraphics[width=\linewidth]{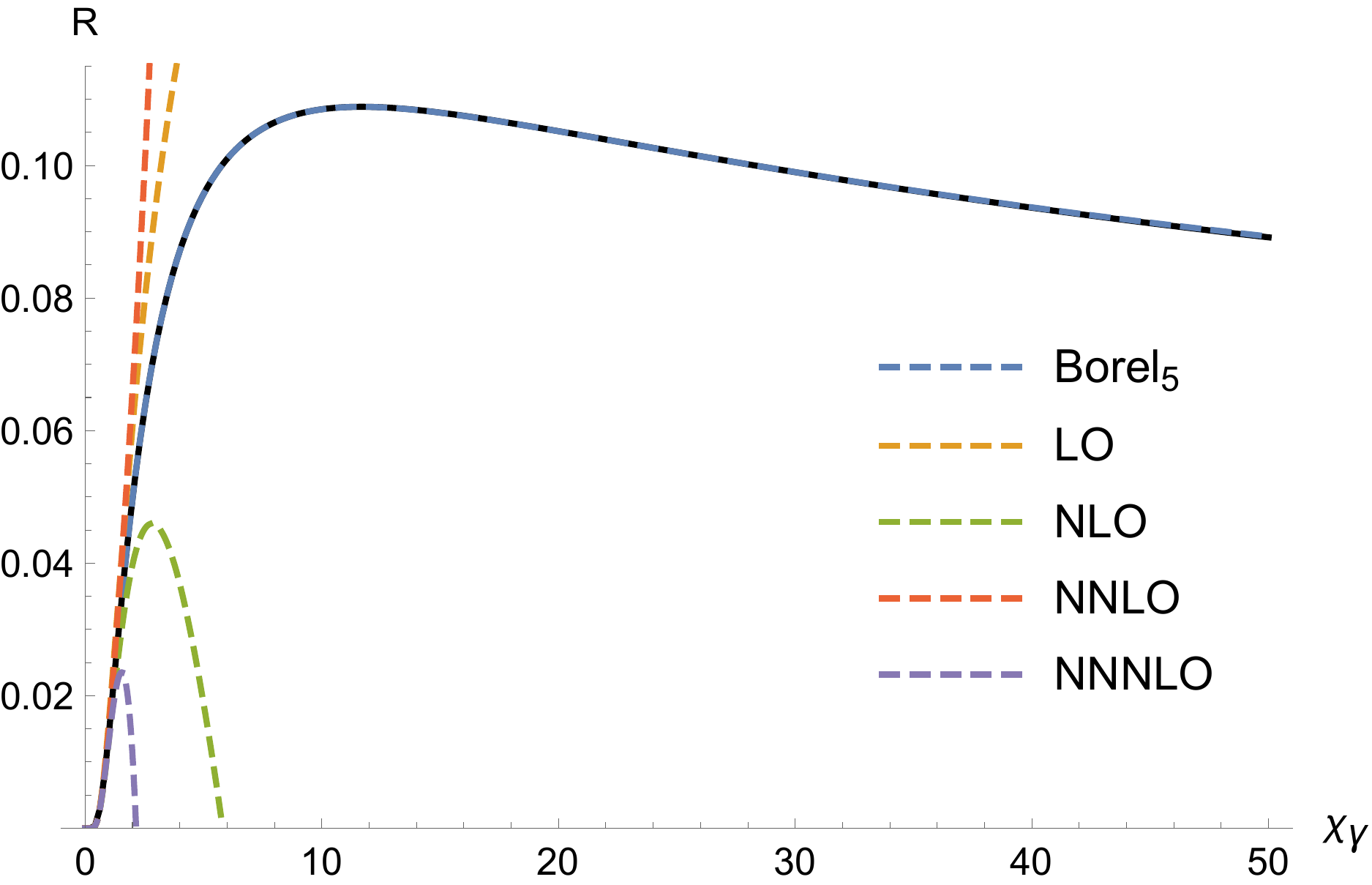}
\includegraphics[width=\linewidth]{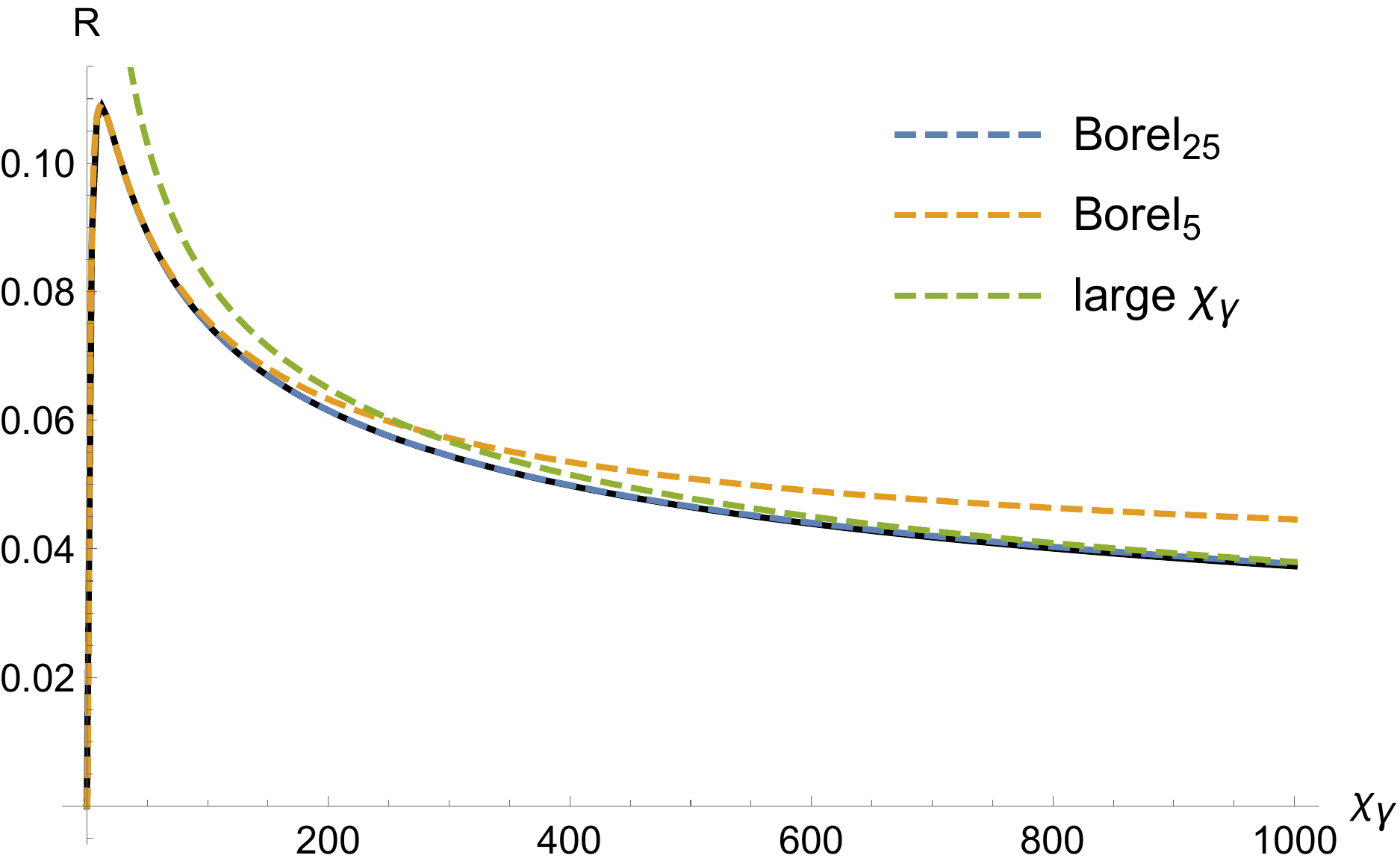}
\caption{$\text{Borel}_{i}$ gives the Pad\'e-Borel resummation with a diagonal Pad\'e approximant with $N=M=i$. LO is the leading small-$\chi$ approximation~\eqref{LCFBWlow}, and NLO, NNLO and NNNLO are obtained by including the first couple of terms in the direct sum of the perturbation series~\eqref{Tdefinition}. The ``large-$\chi_\gamma$'' line shows~\eqref{LCFBWhigh}. The black solid lines show the exact result.}
\label{BWBorelFig}
\end{figure}
  
In Fig.~\ref{BWBorelFig} we compare the direct perturbation series and the resummation with the exact result. We see that, at sufficiently small $\chi_\gamma$, the leading order~\eqref{LCFBWlow} gives a good approximation, but as we increase $\chi_\gamma$ it starts to deviate. Since the power series in $\chi_\gamma$ is a divergent asymptotic series, a direct summation of higher-order terms does not help. However, the Pad\'e-Borel resummed series gives an excellent agreement with the exact result already at Pad\'e order $M=N=5$. In fact, $N=5$ works up to large $\chi_\gamma$: At $\chi_\gamma=100$, the exact result is $\approx0.0749$, compared to $\approx0.0755$ for the resummed series. Going to $N=25$ we find that the Pad\'e-Borel resummed series has a large overlap with the large-$\chi$ approximation~\eqref{LCFBWhigh}. This is what we wanted to see; the small-$\chi_\gamma$ expansion gives divergent power series, but by resumming this series with Borel-Pad\'e methods we obtain a good approximation all the way up to the region where the leading large-$\chi_\gamma$ approximation becomes good.  

In fact, we can obtain an even better agreement by performing a conformal transformation of the Borel transform before forming a Pad\'e approximant, as described in~\cite{Costin:2019xql,Costin:2020hwg}. By numerically matching the ratio of neighboring large-order coefficients of the Borel transform onto the following asymptotic form (cf. Richardson extrapolation in~\cite{Dondi:2020qfj})
\be
\frac{{\rm BT}_n}{{\rm BT}_{n-1}}=c_0+\frac{c_1}{n}+\frac{c_2}{n^2}+\dots \;,
\ee
where ${\rm BT}_n=T_n/n!$, we find that this ratio converges to $c_0=-3/8$. This means that the Borel transform has a finite radius of convergence given by $|t|<8/3$, and a singularity at $t=-8/3$. We therefore replace $t$ in the truncated Borel series with the conformal variable $z$ given by 
\be\label{conformalzt}
z=\frac{\sqrt{1+\frac{3t}{8}}-1}{\sqrt{1+\frac{3t}{8}}+1} \qquad t=\frac{32z}{3(1-z)^2} \;.
\ee   
The next steps are to expand the resulting function in a power series in $z$ to the same order, make a Pad\'e approximant of the new series, and finally express $z$ in terms of $t$. This gives a Pad\'e-conformal resummation ${\rm PCBT}(t)$ of the original truncated Borel series~\cite{Costin:2019xql,Costin:2020hwg}. The final step is to perform the Laplace transform in~\eqref{LaplaceTransform} with ${\rm PCBT}(t)$ instead of ${\rm PBT}(t)$. 

The result is quite impressive: At $\chi_\gamma=10^3$ the relative error $(R_{\rm approx}/R_{\rm exact})-1$ is $\{0.2,0.01,0.05,5\times10^{-7},0.02\}$ for $\{{\rm PB}_5,{\rm PB}_{25},{\rm PCB}_5,{\rm PCB}_{25},\eqref{LCFBWhigh}\}$, where the subscripts $5$ and $25$ stand for the order $N=M$ in the diagonal Pad\'e approximant. We see that by including the conformal step, the relative error at $N=5$ is on the same order of magnitude as $N=25$ for the case without the conformal step. With $N=25$ the conformal approximation gives an extremely high precision, with a relative error several orders of magnitude smaller than the large-$\chi_\gamma$ approximation~\eqref{LCFBWhigh}. At $\chi_\gamma=10^4$ we have a relative error of $\{2\times10^{-4},4\times10^{-3}\}$ for $\{{\rm PCB}_{25},\eqref{LCFBWhigh}\}$, so at such a very large $\chi_\gamma$ the resummation ${\rm PCB}_{25}$ still gives a very high precision and a relative error that is one order of magnitude smaller than the large-$\chi_\gamma$ approximation~\eqref{LCFBWhigh}. At $\chi_\gamma=10^5$ we have a relative error of $\{7\times10^{-3},8\times10^{-4}\}$ for $\{{\rm PCB}_{25},\eqref{LCFBWhigh}\}$, so ${\rm PCB}_{25}$ still gives a relative error of less than one percent. We see that, while the large-$\chi_\gamma$ approximation eventually gives a higher precision, this only happens at a very high $\chi_\gamma$. In fact, this only happens as $\alpha\chi_\gamma^{2/3}$ becomes large, and then one would not trust the leading order in the $\alpha$ expansion anyway. So, 
if we limit ourselves to $\alpha\chi_\gamma^{2/3}$ not large, then the resummation gives a remarkable precision over the entire range of $\chi_\gamma$.


\subsection{Trident}

\begin{figure}
\includegraphics[width=\linewidth]{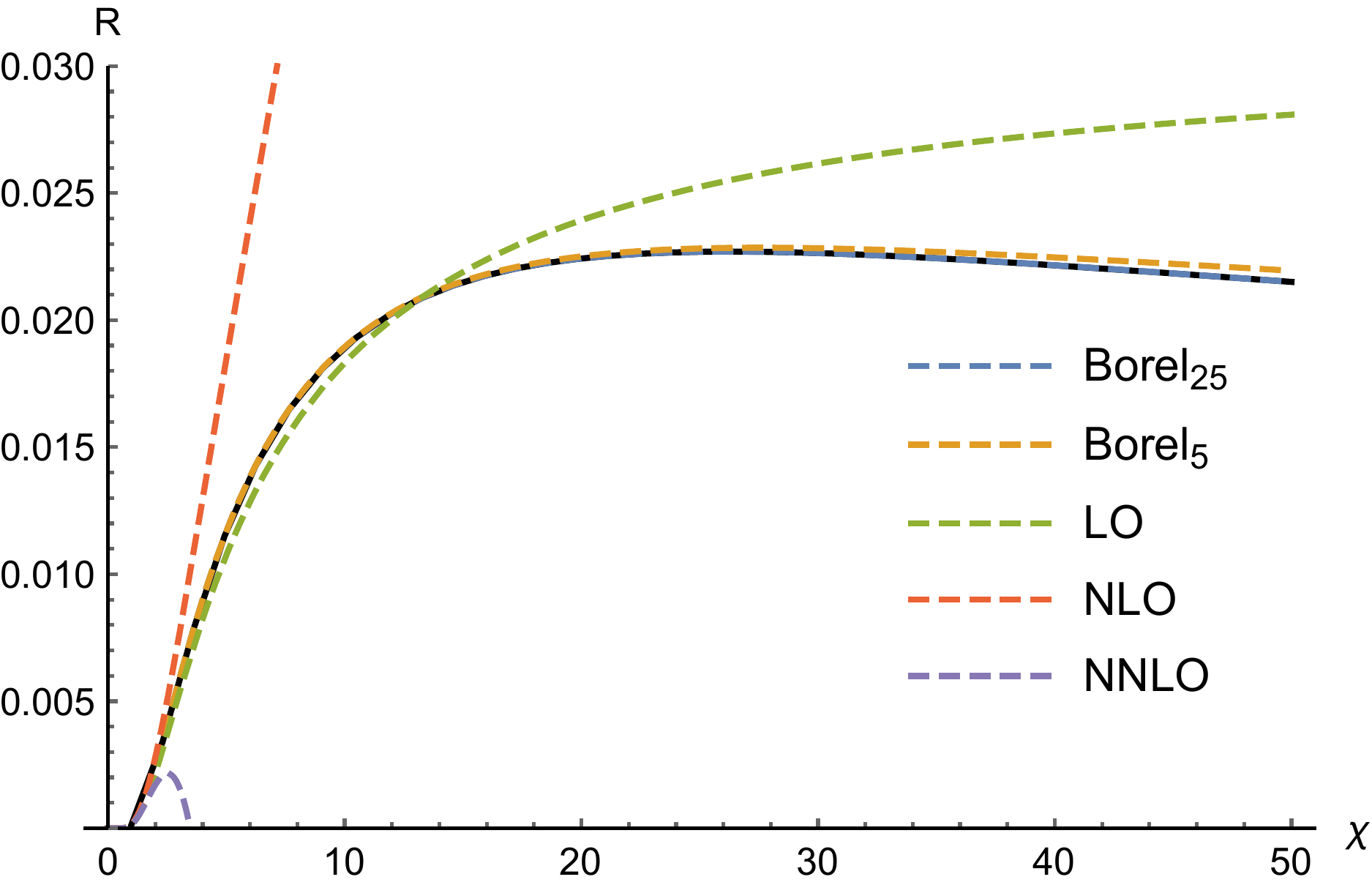}
\includegraphics[width=\linewidth]{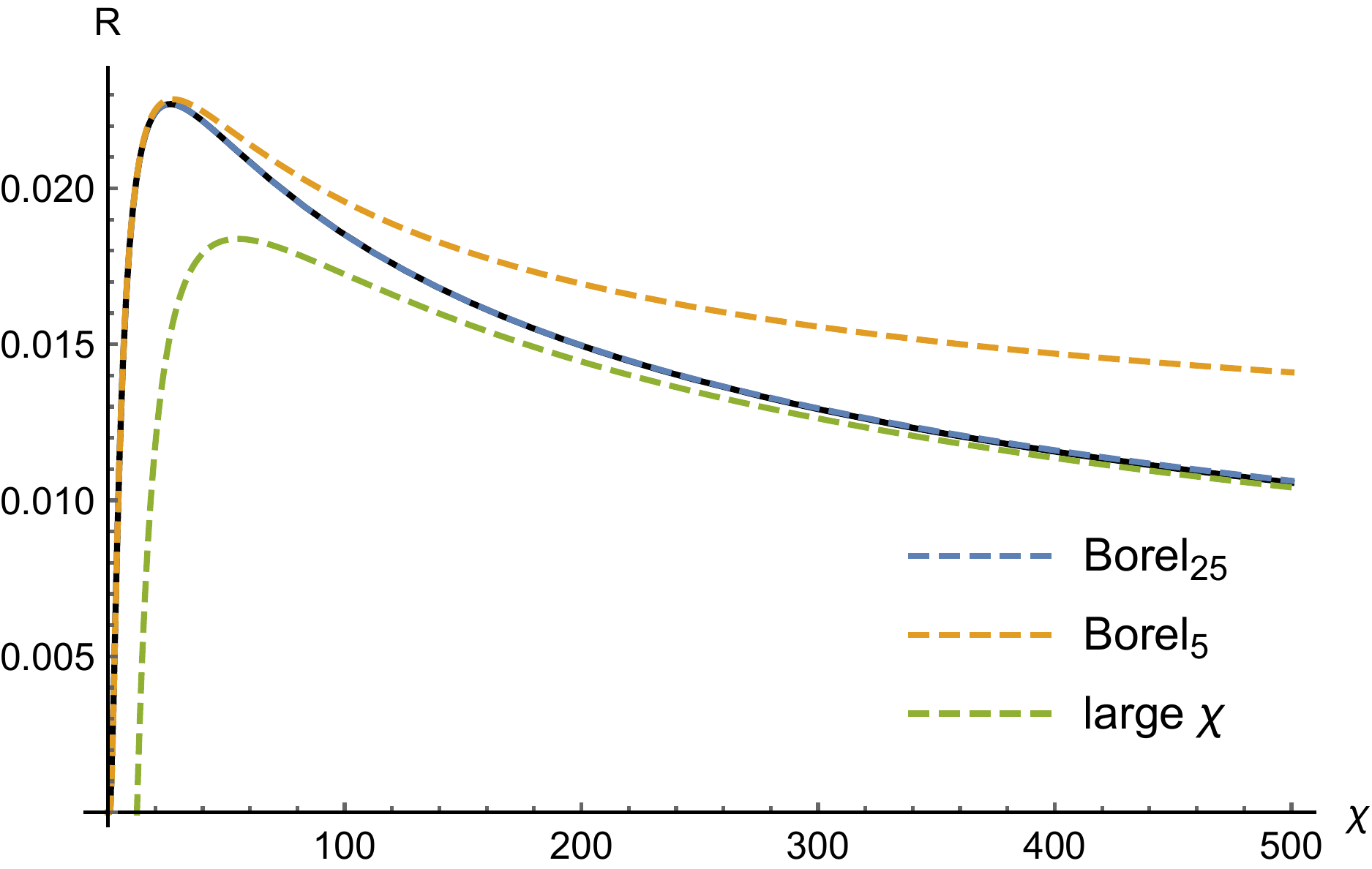}
\caption{Same as Fig.~\ref{BWBorelFig} but for the two-step part of trident.}
\label{tridentBorelFig}
\end{figure}

We will now use the above resummation method for trident. In comparison with previous studies using resummation methods for Schwinger pair production~\cite{Florio:2019hzn,Chadha:1977my,Dunne:1999uy}, note that our expansion parameter $\chi$ gives the field strength in the rest frame of the initial electron in terms of the critical field. 
While the one-step part eventually becomes larger than the two-step part as the energy increases, in this section we will assume that $a_0$ and the pulse length are large enough such that the dominant contribution is given by the two-step part. In LCF this is given by
\be
\mathbb{P}_{\rm two}=\alpha^2a_0^2\int\ud\sigma_{43}\ud\sigma_{21}\theta(\sigma_{43}-\sigma_{21})R \;,
\ee
where (see e.g.~\cite{Dinu:2017uoj,King:2013osa})
\be
\begin{split}
R=&-\int\ud s_1\ud s_2\theta(s_3)\frac{1}{\chi^2q_1^2}\bigg\{\frac{\text{Ai}'(\xi_1)}{\xi_1}\frac{\text{Ai}'(\xi_2)}{\xi_2}+ \\
&\left[\text{Ai}_1(\xi_1)+\kappa_{01}\frac{\text{Ai}'(\xi_1)}{\xi_1}\right]\left[\text{Ai}_2(\xi_2)-\kappa_{23}\frac{\text{Ai}'(\xi_2)}{\xi_2}\right]\bigg\} \;,
\end{split}
\ee
where $\xi_1=(r_1/\chi(\sigma_{21}))^\frac{2}{3}$ and $\xi_2=(r_2/\chi(\sigma_{43}))^\frac{2}{3}$. This expression allows for a locally constant $\chi(\sigma)$, but we will for simplicity consider a constant field. For $\chi\ll1$ we obtain as above
\be
R=\frac{T}{32}\exp\left\{-\frac{16}{3\chi}\right\} \;,
\ee
where
\be
T=1+\frac{31}{216}\chi-\frac{3871}{31104}\chi^2+\frac{492505}{4478976}\chi^3+\dots \;.
\ee
For $\chi\gg1$ we have~\eqref{PtwoLCFchi}. We can again obtain, without much work, the first $\sim70$ terms in $T$. We again find a series with factorially growing coefficients, so we use the Borel-Pad\'e method. The results are shown in Fig.~\ref{tridentBorelFig}. We again find that the Borel-Pad\'e method gives excellent agreement with the exact numerical result up to large values of $\chi$. 

Given the impressive improvement found in the previous section for nonlinear Breit-Wheeler by making a conformal transformation, one would of course also like to make a similar transformation for trident. However, for trident we find that the ratios of neighboring Borel coefficients, ${\rm BT}_n/{\rm BT}_{n-1}$, does not converge (we have calculated $>70$ coefficients). Instead, we find at large $n$ a ratio that goes periodically through 4 different values $\{...,-0.48,-0.23,-0.07,+0.69,-0.48,...\}$. This indicates the presence of complex conjugate pair of singularities on the radius of convergence~\cite{GuttmannVol13}. Compare this with the Breit-Wheeler case, where the only convergence-limiting singularity is on the negative real axis. So, we cannot use the standard ratio test to determine the radius of convergence. One could still use Cauchy-Hadamard's theorem, which gives the radius from $1/r=\underset{n\to\infty}{\text{lim sup }}|{\rm BT}_n|^\frac{1}{n}$, but this converges slowly~\cite{GuttmannVol13}. A better approach is to use Mercer-Robert's procedure~\cite{MercerRoberts}: The radius of convergence is given by $\lim_{n\to\infty}1/B_n$, where
\be
B_n=\left(\frac{{\rm BT}_{n+1}{\rm BT}_{n-1}-{\rm BT}_n^2}{{\rm BT}_n{\rm BT}_{n-2}-{\rm BT}_{n-1}^2}\right)^\frac{1}{2} \;.
\ee      
For the coefficients that we have calculated, $B_n$ has some small oscillations at large $n$, but we find that the radius of convergence is given by $|t|\sim3.8$. The Mercer-Robert's procedure also gives the positions of the conjugate pair of convergence-limiting singularities $re^{\pm i\theta}$, from the $n\to\infty$ limit of
\be
\cos\theta_n=\frac{1}{2}\left(\frac{{\rm BT}_{n-1}B_n}{{\rm BT}_n}+\frac{{\rm BT}_{n+1}}{{\rm BT}_nB_n}\right) \;.
\ee 
We find $\theta=3\pi/4$, so the singularities closest to the origin are at $\sim 3.8e^{\pm3i\pi/4}$. We can confirm this by plotting Pad\'e approximants\footnote{As is well known, Pad\'e approximants can exhibit spurious poles, so we have plotted several different Pad\'e approximants to make sure that any singularity is genuine.} of the Borel transform. The Pad\'e approximants do indeed have singularities at $\sim 3.8e^{\pm3i\pi/4}$. The Pad\'e approximants also show singularities on the real axis at $t<-5$, i.e. further away from the origin\footnote{It could be that $t=-16/3$ is a special point.}.    
The fact that the singularities closest to the origin do not lie on the real axis might suggest using a different type of conformal map, e.g. as in~\cite{Bervillier:2008an}. However, we have found that a conformal map on the form~\eqref{conformalzt}, but with the replacements $3/8\to5$ and $32/3\to20$ (this is motivated by the presence of singularities at $t<-5$), still gives a significant improvement: For $\chi=10^3$ we have a relative error of $\{0.05,6\times10^{-5}\}$ for $\{{\rm PB}_{25},{\rm PCB}_{25}\}$, and for $\chi=10^4$ we have $\{0.3,0.003,0.001\}$ for $\{{\rm PB}_{25},{\rm PCB}_{25},\eqref{PtwoLCFchi}\}$. So, even for $\chi=10^4$ the relative error of this conformal map is on the same order of magnitude as the large-$\chi$ approximation~\eqref{PtwoLCFchi}. Since one might anyway want to keep $\alpha\chi^\frac{2}{3}$ from becoming large, this conformal map seems good enough for our purpose.  

So, we can obtain a good approximation at large $\chi$ by resumming the small-$\chi$ expansion. It is also interesting to note that the small-$\chi$ expansion is obtained by expanding around the saddle point where the three final-state particles have the same longitudinal momentum $s_1=s_2=s_3=1/3$. The expansion coefficients around this point contain the information needed for large $\chi$, even though the spectrum is sharply peaked at $s_1\lesssim1$ at large $\chi$. So, we are expanding around a point at which the value of the spectrum is negligible compared to the spectrum's maximum at large $\chi$.

\subsection{Hypergeometric/Meijer-G resummation}

In this section we will use some resummation methods~\cite{Mera:2014sfa,Mera:2018qte,KleinertHyperResum} which are particularly suitable for functions with a branch cut. The first step is still to calculate the truncated Borel transform~\eqref{BorelBWdefinition}, but then this series is resummed using hypergeometric functions ${}_{q+1}F_q$ instead of the Pad\'e-conformal methods. Assuming that we only have the perturbative information, then the resummation is taken as~\cite{Mera:2018qte}
\be
{\rm HBT}(t)={}_{q+1}F_q\left(\begin{matrix} a,a_1,...,a_q \\ b_1,...,b_q\end{matrix}; \frac{t}{t_0}\right) \;,
\ee
where $a=1$ and the $N=2q+1$ constants $a_i$, $b_i$ and $t_0$ are obtained by expanding the hypergeometric function in a series in $t$ and matching with the Borel transform truncated at $t^N$. In practice, this is conveniently done by~\cite{Mera:2018qte} matching the first $N$ ratios of the Borel coefficients, ${\rm BT}_{n+1}/{\rm BT}_n$ onto a Pad\'e approximant in the variable $n$
\be
\frac{\sum_{i=0}^q p_in^i}{1+\sum_{i=1}^q q_in^i} \;,
\ee
and then comparing with the series-definition of ${}_{q+1}F_q$.
The Laplace transform can then be expressed compactly in terms of a Meijer-G function \cite{MeijerDLMF,LukeBook},
\be\label{MeijerResumT}
\begin{split}
T_{\rm re}^{\rm (M)}(\chi_\gamma)=&\int_0^\infty\frac{\ud t}{\chi_\gamma}e^{-t/\chi_\gamma}{\rm HBT}(t) \\
=&\frac{\prod_{i=1}^q\Gamma(b_i)}{\prod_{i=0}^q\Gamma(a_i)}G^{q+2,1}_{q+1,q+2}\left(\begin{matrix} 1,b_1,...,b_q \\ 1,a,a_1,...,a_q\end{matrix};-\frac{t_0}{\chi_\gamma}\right) \;.
\end{split}
\ee 

For nonlinear Breit-Wheeler, this new resummation allows us to obtain a high precision at large $\chi_\gamma$ with much fewer terms than with the Pad\'e-conformal approach. Already at $N=5$ we find a relative error of less than $5\%$ for $\chi_\gamma=10^3$. At $N=7$ we have encountered some instabilities that seem to be related to the fact that one $a_i$ is very close to one $b_i$, i.e. we are close to a point where the hypergeometric and Meijer-G functions are reduced to a lower order. However, at $N=9$ we find a relative error of less than $3\times10^{-3}$ at $\chi_\gamma=10^3$, and less than $8\times10^{-3}$ at $\chi_\gamma=10^4$.

This is already an impressive improvement, but one can obtain a high precision with even fewer terms by using other known facts in addition to the perturbative data~\cite{KleinertPhi4,KleinertHyperResum,Shalaby:2018unq}. In our case, we know that the Borel transform has a singularity at $t=-8/3$, which can be used to set $t_0=-8/3$. We could then let $a$ be a constant to be obtained by matching in the same way as the other $a_i$ and $b_i$. However, we also know the asymptotic scaling at large $\chi_\gamma$~\eqref{LCFBWhigh}, $R\sim1/\chi_\gamma^\frac{1}{3}$, which, together with the asymptotic limit of the Meijer-G function, can be used to fix one of the constants, e.g. $a=1/3$ (the other constants will be larger so that $a=1/3$ gives the leading asymptotic scaling). $2q+1$ terms are now needed to fix the constants in ${}_{q+1}F_q$ and the overall prefactor. Already at $q=1$, i.e. with only 3 terms in the perturbation series, we find a relative error of $0.01$ at $\chi_\gamma=10^3$, and $0.011$ at $\chi_\gamma=10^4$. At $q=2$ we again encounter an instability.     At $q=3$ we find a relative error of $1.9\times10^{-3}$ at $\chi_\gamma=10^3$, and $2.5\times10^{-3}$ at $\chi_\gamma=10^4$. And at $q=4$ we find a relative error of $1.5\times10^{-4}$ at $\chi_\gamma=10^3$, and $2.0\times10^{-4}$ at $\chi_\gamma=10^4$.

Thus, the new hypergeometric/Meijer-G resummation methods allow for a high precision up to very large $\chi_\gamma$ with relatively few terms from the perturbation series. However, this does not mean that we can forget about the Pad\'e-conformal methods: The hypergeometric/Meijer-G resummation is particularly suitable for functions with a branch cut, but for trident we saw above that the Borel transform has a more complicated structure, with the radius of convergence limited by a complex-conjugate pair of singularities rather than one singularity on the negative axis. So, it is not a priori clear that the hypergeometric/Meijer-G resummation would work for the trident case. We have nevertheless tried it and found that with $N=3$ (using only the perturbative data) the resummation is good up to $\chi_\gamma\sim20$. However, because of the instabilities mentioned above, we have not been able to extend this by increasing $N$.  
One could try take the second line in~\eqref{MeijerResumT} as an ansatz and fix some of the constants by matching with the large-$\chi$ scaling~\eqref{PtwoLCFchi}, which might work since for e.g. $a_1-a_2=0$ (or an integer) the large-$\chi$ limit of involves log terms (cf.~\cite{Shalaby:2020vpm}). 
However, we leave this to future studies. In the next section we will instead consider another new resummation method.

\subsection{Confluent hypergeometric resummation}

In this section we will use the resummation method introduced in~\cite{Alvarez:2017sza}. It is similar to the usual Borel-Pad\'e method, but allows us to use the large-$\chi$ scaling to improve the convergence. In this approach the resummed $T$ is given by
\be\label{TASdefinition}
T_{\rm re}^{\rm AS}(\chi)=\sum_{i=1}^{n}\frac{c_i}{-\chi_i}\phi\left(-\frac{\chi}{\chi_i}\right) \;,
\ee 
where $\phi$ is some suitably chosen special function and the $2n$ constants $c_i$ and $\chi_i$ are determined by matching the $\chi$-series expansion of the two sides. This requires the first $2n$ coefficients of $T$. The following function was proposed in~\cite{Alvarez:2017sza},
\be
\phi(z)=z^{-a}U\left(a,1+a-b,\frac{1}{z}\right) \;,
\ee 
where $U$ is the confluent hypergeometric function. A simple way~\cite{Alvarez:2017sza} to find the the constants in~\eqref{TASdefinition} is to first calculate the $[n-1,n]$ Pad\'e approximant of
\be
\sum_{i=0}^{2n-1}\frac{T_i}{\phi_i}\chi^i \;,
\ee
where $\phi_k=(a)_k(b)_k/k!$, 
and then $\chi_i$ are given by the poles of this approximant (these are simple poles in our case) and $r_i$ are the corresponding residues. $a=b=1$ gives the usual Pad\'e-Borel resummation, but $a$ and $b$ can be chosen such that the large-$\chi$ limit of $\phi(\chi)$ behaves as the known limit of $T$.

In the Breit-Wheeler case, we can take $a=1/3$ and $b=1$, for which $\phi$ can be expressed in terms of an incomplete gamma function with asymptotic scaling $\phi(\chi)\sim1/\chi^{1/3}$, just as in~\eqref{LCFBWhigh}. This choice leads to a resummation that seems competitive with the Meijer-G resummation: At $n=1$, i.e. with only the first two terms in $T$, the relative error at $\chi=10^3$ and $10^4$ is $\sim0.02$; at $n=4$ the relative error is $\sim10^{-3}$ at $\chi=10^3$ and $10^4$. Note that the relative error is only slightly larger at $\chi\sim10^4$ compared to $\chi\sim10^3$ because the large-$\chi$ scaling is built into the resummation function $\phi$. An advantage with this resummation is that it seem relatively fast. Another advantage is that we can also use it for trident.

In the trident case, we take $a=b=2/3$, which gives a large-$\chi$ limit $\phi(\chi)\sim[\ln(\chi)+\text{const.}]/\chi^{2/3}$, which matches the scaling of the leading term in~\eqref{PtwoLCFchi} (but not $\text{const.}$). In this case we find that we need larger $n$ compared to the Breit-Wheeler case. $n=1$ gives nonsensical results. At $n=2$ we find a relative error of $\{0.051,0.066\}$ at $\chi=\{10^3,10^4\}$. At $n=10$ the relative error is $\{9.7\times10^{-4},1.5\times10^{-3}\}$ for $\chi=\{10^3,10^4\}$. The relative error seems to decrease quite slowly as one increases $n$ (also in the Breit-Wheeler case). However, this resummation still requires much fewer terms than the Borel-Pad\'e-conformal method. We leave it to future studies to determine whether a significant improvement can be obtained by choosing a different $\phi$ (maybe a superposition of two $U$) in order to match both the $\ln(\chi)/\chi^{2/3}$ and the $1/\chi^{2/3}$ part of~\eqref{PtwoLCFchi}.  
 
So, we have seen that the resummation methods introduced in~\cite{Alvarez:2017sza} gives a significant improvement over the standard Borel-Pad\'e-conformal method, both in the Breit-Wheeler and the trident case. This still does not mean that one can forget about the Borel-Pad\'e-conformal method, because in some cases the large-$\chi$ (or equivalent) limit might be unknown\footnote{However, even if the large-$\chi$ limit is unknown, it could still be useful to make the large-$\chi$ scaling of the basis function explicit and then vary it until the best convergence is reached, as described in~\cite{KleinertPhi4}.}.

\section{Conclusions}\label{conclusionsSection}

We have obtained new high-energy approximations in the regime where the energy parameter $b_0$ is the largest parameter and where the direct part of the one-step dominates over the two-step. Our high-energy approximation interpolates between the old literature result in the perturbative limit $a_0\ll1$ and previous result for pair production by the superposition of a Coulomb field and a constant-crossed field in the $a_0\gg1$ limit. In between, for arbitrary $a_0$, we find that the high-energy approximation of trident coincides with pair production by the superposition of a Coulomb field and a general, inhomogeneous plane wave. 

Our high-energy approximation is the sum of a logarithmic term, $\ln b_0$, and a $b_0$-independent term. We find that the logarithmic term can be obtained with a Weizs\"acker-Williams equivalent photon approximation. Taking $a_0$ large usually means that the field can be treated as locally constant. However, taking first the energy parameter $b_0$ to be the largest parameter (our new approximation) and then taking $a_0$ large does not commute with first taking $a_0$ to be the largest parameter (standard LCF) and then taking $b_0$ (or $\chi$) large. So, the fact that our new high-energy approximation agrees with the standard Weizs\"acker-Williams approximation (to leading logarithmic order) explains why the latter does not agree with previous LCF results.   

Another interesting difference from the LCF regime is that, while the leading order in the large-$a_0$ limit of the large-$b_0$ approximation is local (similar to the LCF regime), the next-to-leading-order correction is nonlocal, i.e. it is given by an integral where the two lightfront-time variables are not forced to be close but can be far apart. In fact, an important contribution to this correction for compact fields comes from the region where both lightfront-time variables are outside the field but on opposite sides. This is a signal that the formation length is longer in the high-energy limit compared to the large-$a_0$ limit. 

We have also showed that in the case where the initial particle is much heavier than the pair, the dominant contribution in the low-energy limit is given by the term in the amplitude that comes from the instantaneous part of the lightfront Hamiltonian.

We have used Borel, Pad\'e and conformal methods to resum perturbation series in $a_0$, $1/a_0$ and $\chi$. The use of Pad\'e approximants for analytical continuation of perturbation series beyond their radius of convergence has a long history in physics. Here we have shown that our new high-energy approximation has a finite radius of convergence in $a_0$, but by forming Pad\'e approximants we can go beyond this radius of convergence and obtain a good agreement with the large-$a_0$ approximation in an interval of intermediate $a_0$ values. By making a conformal transformation before Pad\'e resummation we obtain agreement with the large-$a_0$ approximation up to much larger $a_0$. We have also used a Pad\'e approximant to analytically continue a power series in $1/a_0$ in the low energy (saddle-point) regime. Finally, we have considered the $\chi$ dependence of nonlinear Breit-Wheeler pair production and the two-step part of trident in the LCF regime. At small $\chi$ the probability can be expanded in a power series in $\chi$ times a ``Schwinger-like'' exponential $e^{-\text{const.}/\chi}$. This power series diverges, so we use a Borel transform to obtain a convergent series. Then we use Pad\'e and conformal methods to analytically continue the truncated Borel transform as described in~\cite{Costin:2019xql,Costin:2020hwg,Florio:2019hzn}. This gives us resummation of the originally divergent $\chi$ series into an approximation that agrees with the exact result up to very large $\chi$, with a significant overlap with the leading large-$\chi$ approximation. We have also showed that newer resummation methods~\cite{Mera:2018qte,Alvarez:2017sza}, which are based on hypergeometric/Meijer-G or confluent hypergeometric functions, can significantly reduce the number of terms that have to be calculated in order to get a certain precision.
It would be interesting to further study these sorts of resummation methods for other strong-field processes and for other fields and parameter regimes.

\acknowledgements
G.~T. thanks Victor Dinu and Sebastian Meuren for inspiring discussions about the high-energy limit, and Burkhard K\"ampfer for discussions about the validity of LCF at $a_0\sim1$, and Gerald Dunne, Anton Ilderton and Ralf Sch\"utzhold for commenting on a draft of this paper. 
G.~T. was supported by the Alexander von Humboldt foundation during the first part of this project.

\end{document}